\documentclass[12pt,fleqn]{article} 
\usepackage{graphicx}
\usepackage{amsfonts}
\newcommand{\R}{\mathbb R}
\renewcommand{\r}{\smallsize \mathbb R}
\renewcommand{\Im}{\mbox{Im }}

\newcommand{\ra}{\rightarrow}
\newcommand{\bra}{\langle} \newcommand{\ket}{\rangle}
\newcommand{\be}{\begin{equation}}
\newcommand{\ee}{\end{equation}}
\newcommand{\bea}{\begin{eqnarray}}
\newcommand{\eea}{\end{eqnarray}}
\newcommand{\eps}{\epsilon}

\newcommand{\ffi}{\varphi}

\newcommand{\ep}{\qquad {\vrule height 10pt width 8pt depth 0pt}}

\newcommand{\grintl}{[\kern-.18em [}
\newcommand{\grintr}{]\kern-.18em ]}
\newcommand{\ds}{\displaystyle}

\newtheorem{theorem}{Theorem}[section]
\newtheorem{thm}[theorem]{Theorem}

\newtheorem{lem}[theorem]{Lemma}
\newtheorem{cor}[theorem]{Corollary}
\newtheorem{prop}[theorem]{Proposition}

\def\smallR{\hbox{\scriptsize I\kern-.23em{R}}}
\def\R{\hbox{$\mit I$\kern-.33em$\mit R$}}
\def\C{\hbox{$\mit I$\kern-.6em$\mit C$}}
\def\un{\hbox{$\mit I$\kern-.77em$\mit I$}}
\def\0{\hbox{$\mit I$\kern-.70em$\mit O$}}
\def\r{I\kern-.277em R}

\def\N{\mbox{\bf N}}

\begin{document}

\title{A Mathematical Theory for Vibrational Levels Associated 
with Hydrogen Bonds\\ I\,:\quad The Symmetric Case}

\author{George A. Hagedorn\thanks{Partially
Supported by National Science Foundation
Grants DMS--0303586 and DMS--0600944.}\\
Department of Mathematics and\\
Center for Statistical Mechanics and Mathematical Physics\\
Virginia Polytechnic Institute and State University\\
Blacksburg, Virginia 24061-0123, U.S.A.\\[15pt]
\and
Alain Joye\\
Institut Fourier\\ Unit\'e Mixte de Recherche CNRS-UJF 5582\\
Universit\'e de Grenoble I,\quad
BP 74\\
F--38402 Saint Martin d'H\`eres Cedex, France\\
and\\
Laboratoire de Physique et Mod\'elisation des Milieux Condens\'es\\
UMR CNRS-UJF 5493, Universit\'e de Grenoble I, BP 166\\
38042 Grenoble, France}

\maketitle

\vskip 6mm
\begin{abstract}
We propose an alternative to the usual time--independent Born--Oppenheimer
approximation that is specifically designed to describe molecules with
symmetrical Hydrogen bonds.
In our approach, the masses of the Hydrogen nuclei are scaled
differently from those of the heavier nuclei, and we employ
a specialized form for the electron energy level surface.
Consequently, anharmonic effects play a role in the leading order
calculations of vibrational levels.

Although we develop a general theory, our analysis is
motivated by an examination of symmetric bihalide ions,
such as $FHF^-$ or $ClHCl^-$.
We describe our approach for the $FHF^-$ ion in detail.
\end{abstract}

\newpage
\baselineskip=20pt 

\section{Introduction}
\setcounter{equation}{0}

In standard Born--Oppenheimer approximations, the masses of the electrons
are held fixed, and the masses of the nuclei
are all assumed to be proportional to $\eps^{-4}$.
Approximate solutions to the molecular Schr\"odinger equation are then
sought as expansions in powers of $\eps$.
For the time--independent problem, the electron energy level surface
is also assumed to behave asymptotically like a quadratic function
of the nuclear variables near a local minimum.

In this paper and in a future one \cite{hagjoy11},
we propose an alternative approximation for molecules that contain Hydrogen
atoms as well as some heavier atoms, such as Carbon, Nitrogen, or Oxygen.
Our motivation is to develop an approach that is specifically tailored
to describe the phenomenon of Hydrogen bonding.

In this paper, we examine the specific case of systems with 
symmetric Hydrogen bonds, such as $FHF^-$. In \cite{hagjoy11}, we plan 
to study non--symmetric cases, where the structure of the 
typical electron energy surface is very different. The mathematical analysis
of that situation is consequently completely different.

The model we present here
differs from the usual Born--Oppenhimer model in two ways:
\begin{enumerate}
\item We scale the masses of the Hydrogen nuclei
as $\eps^{-3}$ instead of $\eps^{-4}$.
This is physically appropriate.
If the mass of an electron is $1$, and we define $\eps^{-4}$
to be the mass of a $C^{12}$ nucleus, then $\eps=0.0821$, and
the mass of a $H^1$ nucleus is $1.015\,\eps^{-3}$. 
\item We do not assume that the electron energy level is
well approximated by an $\eps$--independent quadratic function near a
local minimum. Instead, we allow it to depend on $\eps$ and to take a
particular form that we specify below. The particular form we have
chosen is motivated by a detailed examination of the
lowest electronic potential energy surfaces for $FHF^-$ and $ClHCl^-$.
\end{enumerate}

Although symmetric bihalide ions are quite special,
our approach is flexible enough to describe more general phenomena.
For example, the lowest electron energy surface for
$FHF^-$ has a single minimum with the Hydrogen nucleus
mid--way between the two Fluorines. Our model can handle
situations with single or double wells in the coordinates
for a Hydrogen nucleus that paticipates in Hydrogen bonding. 
We hope that the ideas in this paper and \cite{hagjoy11}
might provide some insight into some properties of Hydrogen bonded systems.

Our model leads to a different expansion from the usual Born--Oppenheimer
approximation. For Hydrogen nuclei not involved in Hydrogen bonding, the
vibrational energies are of order $\eps^{3/2}$, while the vibrational energies
for the other nuclei and the Hydrogen nuclei involved in the 
symmetric Hydrogen bonding
are of order $\eps^2$. Furthermore, anharmonic effects must be taken into
account for a Hydrogen nucleus involved in
Hydrogen bonding at their leading order, $\eps^2$.
In the standard Born--Oppenheimer model, all vibrational energies appear
in a harmonic approximation at order $\eps^2$.
Anharmonic corrections enter at order $\eps^4$.

\vskip 5mm
We present our ideas only in the simplest possible situation.
In that situation, there are only 3 nuclei, and they are constrained
to move along a fixed line. We plan to study more general possibilities,
such as bending of the molecule, in the future.

\vskip 5mm
The paper is organized as follows: In Section \ref{simplest}, we present
the formal expansion. In Section \ref{MathSection} we state our rigorous
results as Theorems \ref{lead} and \ref{asexp}. The proofs of some
technical results are presented in Section \ref{tech}.

\vskip 5mm
\noindent
{\bf Acknowledgements}\quad The authors would like to thank Thierry Gallay
for several helpful conversations. George Hagedorn would like to thank
T. Daniel Crawford for teaching him to use the Gaussian software of
computational chemistry.


\vskip 5mm
\section{Description of the Model}\label{simplest}
\setcounter{equation}{0}

We study a triatomic system with two identical heavy nuclei $A$ and $B$,
and one light (Hydrogen) nucleus $C$. 
We begin by describing the Hamiltonian for this system in
Jacobi coordinates. We let $x_A$ and $x_B$ be the
positions of the heavy nuclei, and let $x_C$ be the
position of the light nucleus $C$. We let their masses be
$m_A=m_B$ and $m_C$.
We let $\ds R=\frac{m_Ax_A+m_Bx_B+m_Cx_C}{m_A+m_B+m_C}$
denote the center of mass of all three nuclei,
and let $\ds x_{AB}=\frac{x_A+x_B}{2}$
denote the center of mass of the heavy nuclei. We let
$W=x_B-x_A$ be the vector from nucleus $A$ to nucleus $B$ and let 
$Z=x_C-x_{AB}$ be the vector from the center of mass of $A$ and $B$ to
$C$. We assume the electronic
Hamiltonian $h_e$ only depends on the vectors between the nuclei,
and we set $m_{AB}=m_A+m_B$ and $M=m_A+m_B+m_C$.
In the original variables, the Hamiltonian has the form
\bea\nonumber
&&-\,\frac{1}{2\,m_A}\,\Delta_{x_A}\,-\,\frac{1}{2\,m_B}\,\Delta_{x_B}
\,-\,\frac{1}{2\,m_C}\,\Delta_{x_C}\,
+\,h_e(x_B-x_A,\,x_C- x_A,\,x_C-x_B).
\eea
In these Jacobi coordinates, it has the form
\bea\nonumber
-\,\frac{1}{2\,M}\,\Delta_{R}\,-\,\frac{m_{AB}}{2\,m_A\,m_B}\,\Delta_{W}\,
-\,\frac{M}{2\,m_{AB}\,m_C}\,\Delta_{Z}\,
+\,h_e(W,\,Z+W/2,\,Z-W/2).
\eea
Since we are interested in bound states, we
discard the kinetic energy of the center of mass. We take the
electron mass to be 1, and the masses of the heavy nuclei to be
$m_A=m_B=\eps^{-4}\mu$, for some fixed $\mu$.
The mass of the light nucleus is
$m_C= \eps^{-3}\nu$, for some fixed $\nu$.
The electronic Hamiltonian $h_e$ then becomes
$h_e(W,Z+W/2,Z-W/2)\equiv h(W,Z)$,
so that the Hamiltonian of interest is
\bea\nonumber
-\,\frac{\eps^{4}}{\mu}\,\Delta_{W}\,-\,
\frac{\eps^{3}}{2\,\nu}\,\left(\,1+\frac{\eps\,\nu}{2\,\mu}\,\right)
\,\Delta_{Z}\,+\,h(W,\,Z).
\eea
This computation is exact and valid in any dimension. 

To simplify the exposition,
we drop the term $\ds \frac{\eps\,\nu}{2\,\mu}$
in the factor that multiplies $\Delta_Z$.
It gives rise to uninteresting, regular perturbation corrections.
Also, for simplicity, we assume $\mu=2$ and $\nu=1$.
This can always be accomplished by trivial rescalings of $W$ and $Z$.

To describe our ideas in the simplest situation,
we restrict $W$ and $Z$ to one dimension. Thus, we are not
allowing rotations or bending of the molecule.
Furthermore, we introduce $\eps$ dependence
of the electronic Hamiltonian to model the pecularities
of symmetric Hydrogen bonds that we describe below.

These considerations lead us to study the Hamiltonian
\be\label{H1}
H_1(\eps)\ =\
-\ \frac{\eps^4}{2}\ \frac{\partial^2\phantom{i}}{\partial W^2}\
-\ \frac{\eps^3}{2}\ \frac{\partial^2\phantom{i}}{\partial Z^2}\
+\ h(\eps,\,W,\,Z).
\ee
The electron Hamiltonian $h(\eps,\,W,\,Z)$ is an operator in the
electronic Hilbert space that depends parametrically on $(\eps,\,W,\,Z)$
and includes the nuclear repulsion terms. For convenience, we assume 
that $h(\eps,\,W,\,Z)$ is a real symmetric operator.

We now describe the specific $\eps$ dependence of
$h(\eps,\,W,\,Z)$ that we assume. 
Although the electron Hamiltonian does not depend on nuclear masses,
the parameter $\eps$ is dimensionless, and thus may play more than
one role.
The dependence of $h$ on $\eps$ we allow is motivated by the
smallness of a particular Taylor series coefficient we observed
in numerical computations for the ground state electron energy
level for the real system $FHF^-$.
We allow only the ground state {\em eigenvalue} to depend on $\eps$.
Otherwise, our electron Hamiltonian is $\eps$--independent.
With the physical value of $\eps$ inserted in our Hamiltonian,
we obtain the true physical Hamiltonian.

From numerical computations of $E(W,\,Z)$ for $FHF^-$,
we observed that the
$Z^2$ coefficient in the Taylor expansion about the
minimum $(W_0,\,0)$ of the ground state potential energy surface
had a small numerial value, on the order of the value of
$\eps=\eps_0$, where $\eps_0$ was defined 
by setting $\eps_0^{-4}$ equal to the nuclear mass of 
the $C^{12}$ isotope of Carbon.

The value of $\eps_0$ is roughly 0.0821.
We define $a_2$ so that the true $Z^2$ Taylor series
term is $a_2\,\eps_0\,Z^2$.
We then obtain $h(\eps,\,W,\,Z)$ by adding
$(\eps-\eps_0)\,a_2\,Z^2$ to the ground state eigenvalue $E(W,\,Z)$.
We make no other alterations to the electron Hamiltonian.
When $\eps=\eps_0$, our $h(\eps,\,W,\,Z)$ equals the
true physical electron Hamiltonian $h(\eps_0,\,W,\,Z)$.

Thus, we assume the ground state electron level has the specific form
\be\label{E1}
\hspace{-1cm}
E_1(\eps,\,W,\,Z)\ =\
E_0\,+\,
a_1\,(W-W_0)^2\,+\,
\bigg(\,a_2\,\eps\,-\,a_3\,(W-W_0)\,\bigg)\,Z^2\,+\,
a_4\,Z^4\,+\ \cdots,
\ee
with $a_j=O(1)$.
As we shall see, the leading order behavior
of the energy and the wave functions
for the molecule are determined from the terms written explicitly
in (\ref{E1}).
The terms not explicitly displayed are 
of orders $(W-W_0)^\alpha\,Z^{2\beta}$, where
$\alpha$ and $\beta$ are non-negative integers that satisfy
$\alpha+\beta\ge 3$.
They play no role to leading order,
but contribute to higher order corrections.

We assume $a_1$, $a_3$, and $a_4$ are positive, but that $a_2$ can be
positive, zero, or negative.
When $a_2$ is negative,
$E_1(\eps,\,W,\,Z)$ has a closely spaced double well near $(W_0,\,0)$
instead of a single local minimum.

To ensure that the leading part of $E_1(\eps,\,W,\,Z)$, 
$$
\widetilde{E}_1(\eps,\,W,\,Z)\ =\
E_0\,+\,
a_1\,(W-W_0)^2\,+\,
\bigg(\,a_2\,\eps\,-\,a_3\,(W-W_0)\,\bigg)\,Z^2\,+\,
a_4\,Z^4,
$$
is bounded below, we assume that either
\bea\label{hyp1}
&&a_3^2\ <\ 4\,a_1\,a_4,
\\ \nonumber&&\hspace{-1.7cm}\mbox{or}
\\ \label{hyp2}
&&a_3^2\ =\ 4\,a_1\,a_4\qquad\mbox{and}\qquad a_2\ \ge\ 0.
\eea
These conditions are equivalent to the property
$\widetilde{E}_1(\eps,\,W,\,Z)\ge -\,C$ for some $C$,
since we can write
\bea\nonumber
\widetilde{E}_1(\eps,\,W,\,Z)\ =\
a_1\,\left(\,(W-W_0)\,-\,\frac{a_3}{2a_1}\,Z^2\,\right)^2\
+\ \left(\,a_4\,-\,\frac{a_3^2}{4a_1}\,\right)\,Z^4\
+\ a_2\,\eps\,Z^2.
\eea
By rescaling with $w=(W-W_0)/\eps$ and $z=Z/\eps^{1/2}$,
we see that the Hamiltonian
$$
-\ \frac{\eps^4}{2}\ \frac{\partial^2\phantom{i}}{\partial W^2}\
-\ \frac{\eps^3}{2}\ \frac{\partial^2\phantom{i}}{\partial Z^2}\
+\ \widetilde{E}_1(\eps,\,W,\,Z)
$$
is unitarily equivalent to $\eps^2$ times the
$\eps$--independent Normal Form Hamiltonian
\be\label{HNF}
H_{{\mbox{\scriptsize NF}}}\ =\
-\ \frac{1}{2}\ \frac{\partial^2\phantom{i}}{\partial w^2}\
-\ \frac{1}{2}\ \frac{\partial^2\phantom{i}}{\partial z^2}\
+\ E_{{\mbox{\scriptsize NF}}}(w,\,z),
\ee
where
\be\label{ENF}
E_{{\mbox{\scriptsize NF}}}(w,\,z)\ =\
a_1\ w^2\ +\ 
\bigg(\,a_2\ -\ a_3\,w\,\bigg)\ z^2\ +\
a_4\ z^4.
\ee

\vskip 5mm \noindent
{\bf Remark}\quad Although we do not use it, further scaling
shows that $H_{{\mbox{\scriptsize NF}}}$ is essentially
a three--parameter model, since the change of variables
$w=\alpha\,s$, $z=\alpha\,t$, yields
\bea\nonumber
H_{{\mbox{\scriptsize NF}}}\ \simeq\
\alpha^{-2}\,\left(\,-\ \frac{1}{2}\
\frac{\partial^2\phantom{i}}{\partial s^2}\
-\ \frac{1}{2}\ \frac{\partial^2\phantom{i}}{\partial t^2}\
+ \alpha_1\,s^2\,+\,\alpha_2\,t^2\,-\,
\alpha_3\,s\,t^2\,+\,t^4\,\right),
\eea
with
\bea\nonumber
\alpha=a_4^{-1/6},\quad
\alpha_1=\frac{a_1}{a_4^{2/3}},\quad
\alpha_2=\frac{a_2}{a_4^{2/3}},\quad\mbox{and}\quad
\alpha_3=\frac{a_3}{a_4^{5/6}}. 
\eea

\vskip 5mm
Under conditions (\ref{hyp1}) or (\ref{hyp2}),
$H_{{\mbox{\scriptsize NF}}}$ is essentially self-adjoint on 
$C_0^\infty(\R^2)$ and has purely discrete spectrum. 
This last property is easy to verify under
condition (\ref{hyp1}), or condition (\ref{hyp2})
with $a_2>0$, because
$E_{{\mbox{\scriptsize NF}}}(w,\,z)$ tends to 
infinity as $\|\,(w,\,z)\,\|\to\infty$.
When (\ref{hyp2}) is satisfied with $a_2=0$, 
the result is more subtle because $E_{{\mbox{\scriptsize NF}}}(w,\,z)$
attains its minimum value of zero along a parabola in $(w,z)$.
In that case we prove that the spectrum is discrete in
Proposition \ref{persson}.

\vskip 5mm
\noindent{\bf Explicit Computations for ${\bf FHF^-}$}\quad
The expression (\ref{E1}) is clearly special. Our computations
for $FHF^-$ that motivate this expression have roughly the
following values, where distances are measured in Angstroms
and energies are measured in Hartrees:
\bea\nonumber
W_0&=&2.287,\\ \nonumber
E_0&=&-200.215,\\ \nonumber
a_1&=&0.26,\\ \nonumber   
a_2&=&1.22\qquad (\,\mbox{if}\quad \eps\ =\ 0.0821\,),
\\ \nonumber 
a_3&=&1.29,\\ \nonumber   
a_4&=&1.62.
\eea

These results came from fitting the output
from Gaussian 2003 using second order Moller--Plesset
theory with the aug--cc--pvtz basis set.
We observed that the process of fitting the data was
numerically quite unstable, and that condition (\ref{hyp1})
was barely satisfied by these $a_j$.
                                                                                
\vskip 8mm
\centerline{\includegraphics[height=6in,width=4in]{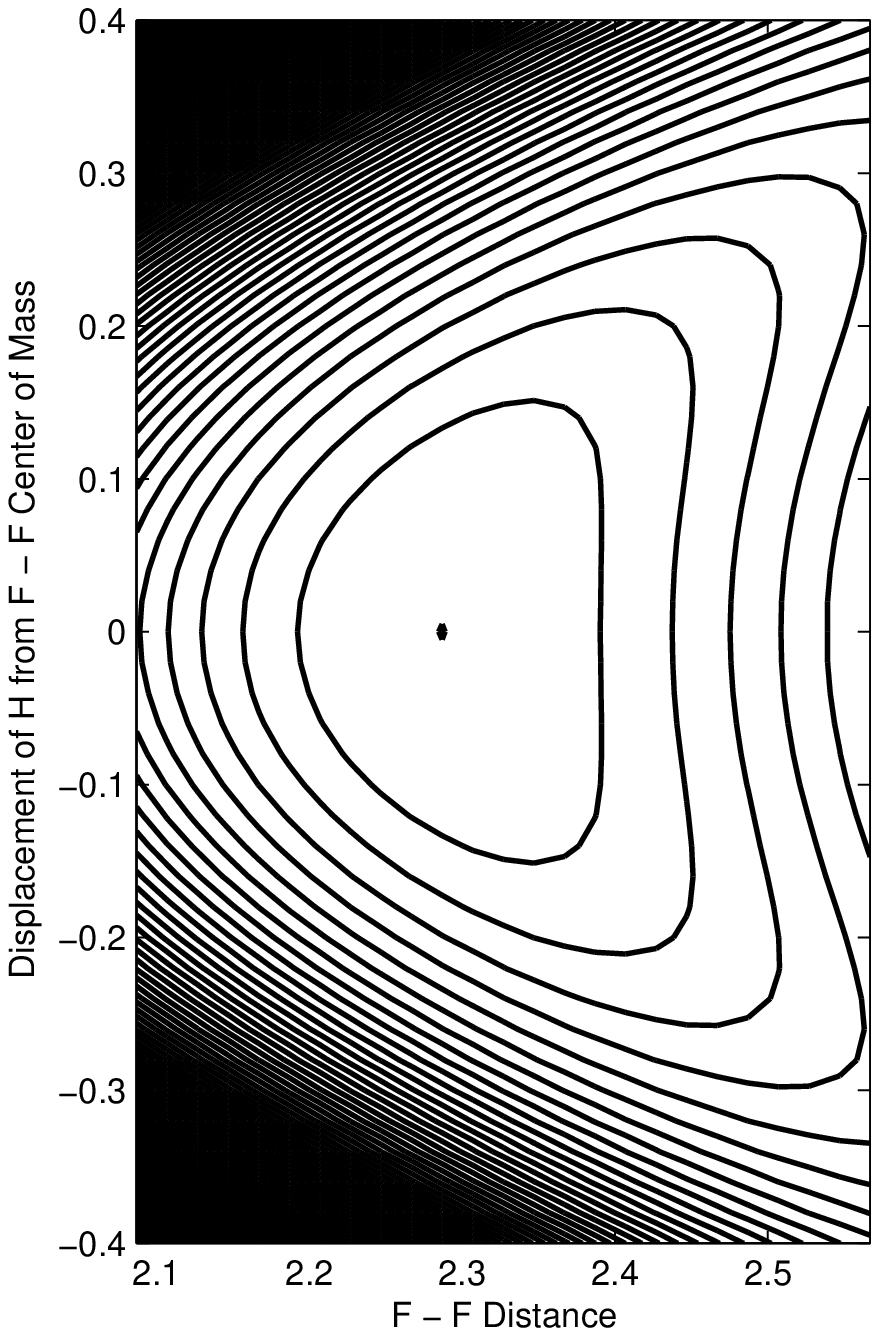}}
                                                                                
\vskip 5mm
\noindent {\bf Figure 1.}\quad Contour plot of the ground state electronic
potential energy surface in the Jacobi coordinates (W,\,Z).
It is obviously not well approximated by a quadratic.
Our technique exploits the flatness of the surface
in the $Z$ direction near the minimum.

\vskip 5mm
The experimentally observed values \cite{EXP} for the
excitation energies to the first symmetric stretching vibrational
mode and the first asymmetric vibrational mode of $FHF^-$ are
$583.05\ \mbox{cm}^{-1}$ and $1331.15\ \mbox{cm}^{-1}$, respectively.
With the values of $a_j$ above, the leading order calculation from our
model predicts $600\ \mbox{cm}^{-1}$ and $1399\ \mbox{cm}^{-1}$.
By leading order, we mean $E_0+\eps^2\,{\cal E}_2$ in the expansion
we present below.
These values depend sensitively on
precisely how we fit the potential energy surface,
which itself depends sensitively on the electron structure calculations.
By comparison, Gaussian 2003 with the aug-cc-pvdz basis set predicts
harmonic frequencies of
$608\ \mbox{cm}^{-1}$ and $1117\ \mbox{cm}^{-1}$. We could not obtain
frequencies for the aug-cc-pvtz basis set from Gaussian because of our
computer limitations.

For some very recent numerical results for vibrational frequencies
of $FHF^-$ that appeared as we were finishing this paper,
see \cite{jcp06}.

\vskip 5mm

We now mimic the technique of \cite{TIBO1} to obtain an expansion
for the solution to the eigenvalue problem for (\ref{H1}).
We could have used the technique of \cite{hagtol2},
but that would have led to more complicated formulas.

For convenience, we replace the variable $W$ by $W-W_0$, so that
henceforth, $W_0=0$.

The technique of \cite{TIBO1} uses the method of multiple scales.
Instead of searching directly for an eigenvector
$\Psi(\eps,\,W,\,Z)$ for (\ref{H1}),
we first search for an eigenvector $\psi(\eps,\,W,\,Z,\,w,\,z)$
for an operator that acts in more variables.
When we have determined $\psi$, we obtain $\Psi$ by setting
$$
\Psi(\eps,\,W,\,Z)\ =\ \psi(\eps,\,W,\,Z,\,W/\eps,\,Z/\eps^{1/2}).
$$
This is motivated physically by the following observation:
The dependence of the electrons on the nuclear coordinates
occurs on the length scale of
$(W,\,Z)$, while the semiclassical quantum
fluctuations of the nuclei occur on the length scale of $(w,\,z)$.
To leading order in $\eps$, these effects behave independently.

\vskip 5mm
The equation for $\psi$ is formally
\be\label{MSequation}
H_2(\eps)\ \psi(\eps,\,W,\,Z,\,w,\,z)\quad =\quad
{\cal E}(\eps)\ \psi(\eps,\,W,\,Z,\,w,\,z),
\ee
where
\bea\nonumber\hspace{-1cm}
H_2(\eps)&=&
-\ \frac{\eps^4}{2}\ \frac{\partial^2\phantom{i}}{\partial W^2}\
-\ \eps^3\,\frac{\partial^2\phantom{i}}{\partial W\,\partial w}\
-\ \frac{\eps^2}{2}\ \frac{\partial^2\phantom{i}}{\partial w^2}\
-\ \frac{\eps^3}{2}\ \frac{\partial^2\phantom{i}}{\partial Z^2}\
-\ \eps^{5/2}\,\frac{\partial^2\phantom{i}}{\partial Z\,\partial z}\
-\ \frac{\eps^2}{2}\ \frac{\partial^2\phantom{i}}{\partial z^2}
\\[4mm] \nonumber
&&+\quad [\,h(\eps,\,W,\,Z)\,-\,E(\eps,\,W,\,Z)\,]\ +\
E(\eps,\,\eps\,w,\,\eps^{1/2}z)
\\[4mm] \label{H2}
&&+\ \,\sum_{m=6}^\infty\ \eps^{m/2}\
\left(\ T_{m/2}(W,\,Z)\,-\,T_{m/2}(\eps\,w,\,\eps^{1/2}z)\ \right).
\eea
The functions $T_{m/2}$ in this expression will be chosen later.
Different choices yield equally valid expansions for $\Psi(\eps,\,W,\,Z)$,
although they alter the expressions for
$\psi(\eps,\,W,\,Z,\,w,\,z)$ by converting $(W,\,Z)$
dependence into $(w,\,z)$ dependence.

\vskip 5mm
In (\ref{H2}), we expand both $E(\eps,\,\eps w,\,\eps^{1/2}z)$ and
$T_{m/2}(\eps w,\,\eps^{1/2}z)$ in Taylor series in powers of $\eps^{1/2}$.
We then make the Ansatz that (\ref{MSequation}) has formal
solutions of the form
\be\label{forsol}\hskip -1cm
\psi(\eps,W,Z,w,z)\ =\ \psi_0(W,Z,w,z)\,+\,
\eps^{1/2}\,\psi_{1/2}(W,Z,w,z)\,+\,
\eps^1\,\psi_1(W,Z,w,z)\,+\,\cdots,
\ee
with
\be\label{expe}
{\cal E}(\eps)\ =\ {\cal E}_0\ +\ \eps^{1/2}\ {\cal E}_{1/2}\ +\ 
\eps^1\ {\cal E}_1\ +\ \cdots.
\ee
We substitute these expressions into (\ref{MSequation}) and
solve the resulting equation order by order in powers of $\eps^{1/2}$.

\noindent
{\bf Note:} The description in this section is purely formal.
In particular, it does not take into account the
cutoffs that are necessary for rigorous results.
The mathematical details are dealt with in the next section.

\vskip 4mm
\noindent\underbar{\bf Order ${\bf 0}$}\quad
The order $\eps^0$ terms require
$$
[\,h(\eps,\,W,\,Z)-E(\eps,\,W,\,Z)\,]\ \psi_0\ +\ E_0\ \psi_0 \ =\
{\cal E}_0\ \psi_0.
$$
We solve this by choosing
$$
{\cal E}_0\ =\ E_0,
$$
and
$$
\psi_0(W,\,Z,\,w,\,z)\ =\ f_0(W,\,Z,\,w,\,z)\ \Phi(W,\,Z),
$$
where $\Phi(W,\,Z,\,\cdot\,)$ is a normalized
ground state eigenvector of $h(\eps,\,W,\,Z)$.
Under our assumptions, we can choose $\Phi(W,\,Z,\,\cdot\,)$ to be real,
smooth in $(W,\,Z)$, and independent of $\eps$.
This choice satisfies
\be\label{connection}
\langle\,\Phi(W,\,Z,\,\cdot\,),\ 
\nabla_{W,Z}\Phi(W,\,Z,\,\cdot\,)\,\rangle_{{\cal H}_{el}}
\ =\ 0,
\ee
where the inner product is in the electronic Hilbert space.
We assume that $f_0(W,\,Z,\,w,\,z)$ is not identically zero.

\vskip 4mm
\noindent\underbar{\bf Order ${\bf 1/2}$}\quad
The order $\eps^{1/2}$ terms require
$$
[\,h(\eps,\,W,\,Z)\,-\,E(\eps,\,W,\,Z)\,]\ \psi_{1/2}\ +\ E_0\ \psi_{1/2}
\ =\
{\cal E}_0\ \psi_{1/2}\ +\
{\cal E}_{1/2}\ \psi_{0}.
$$
The components of this equation in the $\Phi(W,\,Z)$ direction
in the electronic Hilbert space require
$$
{\cal E}_{1/2}\ =\ 0.
$$
The components of the equation orthogonal to $\Phi(W,\,Z)$ 
in the electronic Hilbert space require
$$
[\,h(\eps,\,W,\,Z)\,-\,E(\eps,\,W,\,Z)\,]\ \psi_{1/2}\ =\ 0,
$$
so
$$
\psi_{1/2}(W,\,Z,\,w,\,z)\ =\ f_{1/2}(W,\,Z,\,w,\,z)\ \Phi(W,\,Z).
$$

\vskip 4mm
\noindent\underbar{\bf Orders ${\bf 1}$ and ${\bf 3/2}$}\quad
By similar calculations, the order $\eps^1$ and $\eps^{3/2}$ terms yield
\bea\nonumber
{\cal E}_1&=&{\cal E}_{3/2}\quad =\quad 0,
\\[3mm]\nonumber
\psi_{1}(W,\,Z,\,w,\,z)&=&f_1(W,\,Z,\,w,\,z)\ \Phi(W,\,Z),
\qquad\qquad\mbox{and}
\\[3mm]\nonumber
\psi_{3/2}(W,\,Z,\,w,\,z)&=&f_{3/2}(W,\,Z,\,w,\,z)\ \Phi(W,\,Z).
\eea

\vskip 4mm
\noindent\underbar{\bf Order ${\bf 2}$}\quad
The order $\eps^2$ terms that are multiples of
$\Phi(W,\,Z)$ in the electronic Hilbert space require
\bea\nonumber
&&-\ \frac 12\,\frac{\partial^2f_0}{\partial w^2}(W,\,Z,\,w,\,z)\ -\
\frac 12\ \frac{\partial^2f_0}{\partial z^2}(W,\,Z,\,w,\,z)\ +\
E_{{\mbox{\scriptsize NF}}}(w,\,z)\ f_0(W,\,Z,\,w,\,z)
\\[3mm] \label{parallel2}
&=&{\cal E}_2\ f_0(W,\,Z,\,w,\,z),
\eea
where $E_{{\mbox{\scriptsize NF}}}(w,\,z)$ is given by (\ref{ENF}).

Because of the form of $E_{{\mbox{\scriptsize NF}}}(w,\,z)$,
(\ref{parallel2}) does not separate into two ODE's. 
We do not know ${\cal E}_2$ or $f_0$ exactly, although accurate numerical
approximations can be found easily.
These eigenvalues
and eigenfunctions describe the coupled anharmonic vibrational motion of
all three nuclei in the molecule.
As we commented earlier, hypotheses (\ref{hyp1}) or (\ref{hyp2})
guarantee that the eigenvalues ${\cal E}_2$ are discrete
and bounded below, with normalized bound states $f_0(W,\,Z,\,w,\,z)$
in $(w,\,z)$ for any $(W,\,Z)$.

Later in the expansion, 
we choose the operator $T_3$ so that $f_0$ has no
$(W,\,Z)$ dependence. With this in mind, equation (\ref{parallel2})
determines ${\cal E}_2$ and a normalized function
$f_0(w,\,z)$ (up to a phase)
for any given vibrational level.

\vskip 5mm
The terms of order 2 that are orthogonal to $\Phi(W,\,Z)$ require
$$
[\,h(\eps,\,W,\,Z)\,-\,E(\eps,\,W,\,Z)\,]\ \psi_{2}\ =\ 0,
$$
Thus, 
$$\psi_2\ =\ f_2(W,\,Z,\,w,\,z)\ \Phi(W,\,Z).
$$

We split the scalar functions $f_\alpha(W,\,Z,\,w,\,z)$ with
$\alpha>0$ into two contributions
$$
 \ f_\alpha(W,\,Z,\,w,\,z)\ =\
 f_\alpha^\parallel (W,\,Z,\,w,\,z)\ +\ f_\alpha^\perp (W,\,Z,\,w,\,z)\ 
$$
where for each fixed $W$ and $Z$,
$\ f_\alpha^\parallel (W,\,Z,\cdot,\cdot\,)$ is a multiple of
$f_0(\cdot, \cdot\,)$, and
$\ f_\alpha^\perp (W,\,Z,\cdot,\cdot\,)$ perpendicular to
$f_0(\cdot,\cdot\,)$ in $L^2(\mathbb R^2,\,dw\,dz)$.
Furthermore, we choose the operators $T_{3+m/2}$ later in the expansion
so that $f_\alpha^\parallel (W,\,Z,\cdot,\cdot\,)$ has no $(W,\,Z)$
dependence. We will not precisely normalize our approximate eigenfunctions,
so we henceforth assume $f_\alpha^\parallel(W,\,Z,\,w,\,z)=0$
for all $\alpha>0$.

\vskip 4mm
\noindent\underbar{\bf Order ${\bf m/2}$ with ${\bf m>4}$}\quad
We equate the terms of order $m/2$ and then 
separately examine the projections of the resulting equation
into the $\Phi(W,Z)$ direction in the electron Hilbert space
and into the direction perpendicular to $\Phi(W,Z)$.

From the terms in the $\Phi(W,\,Z)$ direction, we obtain the value
of ${\cal E}_{m/2}$ and an expression for
$f_{(m-4)/2}(W,\,Z,\,w,\,z)=f_{(m-4)/2}^\perp (W,\,Z,\,w,\,z)$.
When $m=6$ we choose $T_3$ so that $f_0$ can be chosen independent
of $(W,\,Z)$. When $m>6$,
we choose $T_{m/2}$,  so that
$f_{(m-6)/2}^\parallel$ can be taken to be zero.

The terms orthogonal to $\Phi(W,\,Z)$ in the electronic Hilbert space
give rise to an equation for
$[\,h(\eps,\,W,\,Z)\,-\,E(\eps,\,W,\,Z)\,]\ \psi_{m/2}$.
This equation has a solution of the form
\bea\nonumber
\psi_{m/2}(W,\,Z,\,w,\,z) &=&
\left(\,f_{m/2}^\parallel (W,\,Z,\,w,\,z)\ +\
f_{m/2}^\perp (W,\,Z,\,w,\,z)\,\right)\ \Phi(W,\,Z)
\\[3mm] \nonumber 
& &+\quad
\psi_{m/2}^{\perp}(W,\,Z,\,w,\,z),
\eea
where $\psi_{m/2}^{\perp}$ is obtained by applying the reduced
resolvent operator
$[\,h(\eps,\,W,\,Z)\,-\,E(\eps,\,W,\,Z)\,]^{-1}_r$
to the right hand side of the equation. 
\\

In the next section, we prove that this procedure yields
a quasimode whose approximate eigenvalue and eigenvector
each have asymptotic expansions to all orders in $\eps^{1/2}$.


\vskip 5mm
\section{Mathematical Considerations}\label{MathSection}
\setcounter{equation}{0}

In this section we present a mathematically rigorous version
of the expansion of Section \ref{simplest}. This involves inserting
cutoffs and proving that many technical conditions are satisfied
at each order of the expansion.

\begin{prop}\label{persson}
Assume (\ref{hyp1}) or (\ref{hyp2}).\\[2mm]
Then, the spectrum of
$\ds H_{NF}\ =\ -\,\frac 12\,\frac{\partial^2}{\partial w^2}\
-\,\frac{1}{2}\,\frac{\partial^2}{\partial z^2}\ +\ E_{NF}(w,z)$ 
is purely discrete. 
\end{prop}

\vskip 5mm\noindent
{\bf Proof}\quad
We use Persson's Theorem (see, {\it e.g.}, \cite{hissig})
to show that the essential spectrum
of $H_{NF}$ is empty. This theorem says 
that if $V\in L^p(\R^n)+L^\infty(\R^n)$, with $p=2$ if $n\leq 3$,
$p>2$ if $n=4$, and $p\geq n/2$ if $n\geq 5$, then the bottom of 
the essential spectrum of $H=-\Delta+V$ is characterized by the 
behavior of the operator at infinity. More precisely,
$$
\inf\ \sigma_{\mbox{\small ess}}(H)\ =\
\sup_{K \in \r^n \atop K \ \mbox{\small compact}}\
\inf_{\ffi\neq 0}\ \left\{\,\frac{\bra\,\ffi,\,H\,\ffi\,\ket}{\|\ffi\|^2}
\ : \
  \ffi\in C^\infty_0(\R^n\setminus K)\,\right\}.
$$
Since $E_{NF}$ does not satisfy the hypotheses,
we replace it with a cut off potential
$$
E_{T}(w,z)=\left\{\matrix{E_{NF}(w,z), & \mbox{if}\ E_{NF}(w,z) \leq
    T,\vspace{2mm}  \cr
T, & \mbox{otherwise}.}\right. 
$$
The operator $H_T\,=\,-\,\Delta+E_T$ is self-adjoint on the domain
of $-\,\Delta$. Because $C_0^\infty$ is a core for both $H_{NF}$
and $H_F$, and 
$\bra\,\ffi,\,H_{NF}\,\ffi\,\ket\ \geq\ \bra\,\ffi,\,H_T\,\ffi\,\ket$,
for any $\ffi\in C_0^\infty$,
the min-max principle shows that 
\be\label{george1}
\inf\ \sigma_{\mbox{\small ess}}(H_{NF})\ \geq\
\inf\ \sigma_{\mbox{\small ess}}(H_T).
\ee

Under hypothesis (\ref{hyp1}) or (\ref{hyp2}) with $a_2>0$, $E_{NF}$
is arbitrarily large for all large arguments. Persson's Theorem easily
shows that $\inf\ \sigma_{\mbox{\small ess}}(H_T)\ =\ T$ for all large
positive $T$. Inequality (\ref{george1}) immediately implies the
proposition.

Thus, we need only consider the case of hypothesis (\ref{hyp2}) with
$a_2=0$. Since $E_T\leq T$, we see that
$\ \inf\ \sigma_{\mbox{\small ess}}(H_T)\ \leq\ T$. We shall use
Persson's Theorem to prove the reverse inequality.

Consider a square $K(R)$ of side $2R>0$, centered at the origin, and
let $\ffi\in C^\infty_0(\R^n\setminus K(R))$. 
We  observe that $E_{NF}(w,\,z)=T$ on the set
$L_R=\{w\leq -R\}\cup \{|z|\geq R, |w|\leq R\}$,
for $R > 2a_1/a_3$, provided $T\leq a_1R^2$. Therefore, 
$$
\int_{L_R}\ \overline{\ffi(w,z)}\
\left(-\partial_w^2/2-\partial_z^2/2+E_{T}(w,z)\right)\,\ffi(w,z)\
dw\,dz\
\geq\ \int_{L_R}\ T\ |\ffi(w,z)|^2\ dw\,dz.
$$
For $w\geq R$, we estimate the integral as follows
\bea
&&\int_{\{w\geq R,\ z\in\r\}}\ \overline{\ffi(w,z)}\
\left(-\partial_w^2/2-\partial_z^2/2+E_{T}(w,z)\right)\,\ffi(w,\,z)\
dw\,dz \nonumber \\[3mm] 
&&\quad\quad\quad\quad\geq\quad\int_{w \geq R}\ 
\int_{z\in\r}\ \overline{\ffi(w,z)}\ 
\left(-\partial_z^2/2+E_{T}(w,z)\right)\,\ffi(w,\,z)\
dz\,dw\nonumber \\[3mm] \label{lobo}
&&\quad\quad\quad\quad\equiv\quad\int_{w\geq R}\ \bra\,\ffi(w,\cdot\,),\,
h_T(w)\,\ffi(w,\cdot\,)\,\ket_{z}\ dw.
\eea
For each value of $w\geq R$, the operator $h_T(w)$ is a
one-dimensional Schr\"odinger
operator with potential given by a (cut off) symmetric quartic double well.
Hence $h_T(w)$ always has a ground state below $T$, for any large $T$.
We shall show that the ground state converges to $T$ as $w\ra \infty$.

To do this, we show that $h_T(w)\ra -\partial_z^2/2+T$ in the
norm resolvent sense. By the resolvent identity, this follows if we 
show that $\|(E_T(w,z)-T)(-\partial_z^2/2+T)^{-1}\|$ tends to zero
as $w\ra\infty$. To show this we use the following claim:\\
There exists $c(T)$, such that, 
$$
\|\,V\,(-\partial_z^2+T)^{-1}\,\|_{L^2(\r)\ra L^2(\r)}\ \leq\
c(T)\ \|\,V\,\|_2,
$$
for all $V\in L^2(\R)$.

By Theorem IX.28 in \cite{rs2},
given any $a>0$, there exists
$b\geq 0$, such that\\
$
\|\ffi \|_\infty\ \leq\ a\,\|(-\partial_z^2)\ffi\|_2\ +\ b\,\|\ffi\|_2,
$\quad
for any $\ffi$ in the domain of $-\partial_z^2$.\quad
This implies that 
\bea
\|\,V\ (-\partial_z^2+T)^{-1}\,\|_{L^2(\r)\ra L^2(\r)}
&\leq &
\|\,V\,\|_{L^\infty(\r)\ra L^2(\r)}\
\|\,(-\partial_z^2+T)^{-1}\,\|_{L^2(\r)\ra L^\infty(\r)}
\nonumber \\ \nonumber
&\leq&c(T)\ \|\,V\,\|_2,
\eea
for some finite $c(T)$. This proves the claim.

In our case, this yields
$$
\|\,(E_T(w,\cdot\,)-T)\,(-\partial_z^2/2+T)^{-1}\,\|
\ \leq\ c(T)\ \|\,(E_T(w,\cdot\,)-T)\,\|_2,
$$
where $c(T)$ is independent of $w$.

Under the assumption that $R\geq \sqrt{T/a_1}$, we have
\bea\nonumber
\|\,(E_T(w,\cdot\,)-T)\,\|_2^2&\leq&
\int_{|w-z^2a_3/(2a_1)|\leq\sqrt{T/a_1}}\ T^2\ dz\\[3mm] \nonumber
&=&2\ T^2\ \sqrt{\frac{2a_1}{a_3}}\ \left(\,\sqrt{w+\sqrt{T/a_1}}\ -\
\sqrt{w-\sqrt{T/a_1}}\,\right) 
\\[3mm] \nonumber
&\leq&T^2\ \sqrt{\frac{2T}{a_3}}\,\ \frac{1}{\sqrt{w-\sqrt{T/a_1}}}
\\[3mm] \nonumber
&\simeq& 
\tilde c(T)\,\ \frac{1}{\sqrt{w}},\quad \mbox{as}\quad w\ra\infty.
\eea
Thus by the resolvent formula and Theorem VIII.19 of \cite{rs1},
$h_T(w)$ converges in norm resolvent sense to $(-\partial_z^2/2+T)$.

We let $P_{\Delta}(H)$ be the
spectral projector on the interval $\Delta\in\R$ for a self-adjoint
operator $H$. Then Theorem VIII.23 of \cite{rs1} implies
that for any positive $a<b<T$,
$$
\lim_{w\ra\infty}\ 
\|\,P_{(a,b)}(h_T(w))\,-\,P_{(a,b)}(-\partial_z^2/2+T)\,\|
\ =\ \lim_{w\ra\infty}\ \|P_{(a,b)}(h_T(w))\|\ =\ 0,
$$
since $\sigma(-\partial_z^2/2+T)\,=\,[T,\infty)$. 
Therefore, the bottom of the spectrum of $h_T(w)$ satisfies 
$$
T\ \ge\ b_T(w)\ \equiv\ \inf\ \sigma(h_T(w))\ \ra\ T
\quad\mbox{as}\quad w\ra\infty.
$$
Using this in (\ref{lobo}), we obtain
\bea\nonumber
&&\int_{\{w\geq R,\ z\in\r\}}\
\overline{\ffi(w,z)}\
(-\partial_w^2/2-\partial_z^2/2+E_{T}(w,z))\,\ffi(w,z)\
dw\,dz\\[3mm] \nonumber
&&\quad\quad\quad\quad\geq\ \inf_{v\geq R}\ b_T(v)\
\int_{\{w\geq R, z\in \r \}}\  |\ffi(w,z)|^2\
dw\,dz.
\eea
Combining all the estimates, we see that for any
$\ffi \in C^\infty_0(\R^n\setminus K(R))$
$$
\bra\,\ffi,\ H_T\ \ffi\,\ket\ \geq\ \inf_{v\geq R}\ b_T(v)\
\|\,\ffi\,\|^2, 
$$
provided $a_1\,R^2\,\geq\,T$. Thus,
$$
\inf\ \sigma_{\mbox{\small ess}}(H_T)\ \geq\
\sup_{R\geq\sqrt{\frac{T}{a_1}}}\ \inf_{v\geq R}\ b_T(v)\ =\ T.
$$
Since $T$ is arbitrarily large, this implies the proposition.
\hfill\ep 

\vskip 5mm
In the usual Born--Oppenheimer approximation, the semiclassical
expansion for the nuclei is based on Harmonic oscillator
eigenfunctions. They have many well-known properties.
Our expansion relies on the analogous properties for
eigenfunctions of $H_{NF}$. The following proposition
establishes some of the properties we need in an
even more general setting.

\vskip 5mm
\begin{prop}\label{eigv}
Let $V$ be a non-negative polynomial, such that
$H=-\Delta + V $ has purely discrete spectrum. 
Let $\ffi(x)$ be an eigenvector of $H$, {\rm i.e.},
an $L^2(\mathbb R^n)$ solution 
of $H\ffi={\cal E}\ffi$, where ${\cal E} > 0$. 
Then, $\ffi \in C^\infty(\R^n)$ 
and  $\nabla\ffi \in L^2(\R^n)$. Moreover, for any $a>0$,
$$
\ffi\in D(e^{a\bra x\ket}),\quad\nabla\ffi\in D(e^{a\bra x\ket}),
\quad\mbox{and}\quad\Delta\ffi\in D(e^{a\bra x\ket}),
$$
where $\ds\bra x\ket =\sqrt{1\,+\,\sum_{j=1}^n\,x_j^2\,}\,$,\quad
and $D(e^{a\bra x\ket})$ denotes the domain of multiplication by
$e^{a\bra x\ket}$. 
\end{prop}

\noindent
{\bf Proof}\quad
Since $V\in C^\infty$, elliptic regularity arguments
(see {\it e.g.}, \cite{h}, Thm 7.4.1) show that all eigenfunctions
are $C^\infty$.

We first show that the $\nabla\ffi$ is $L^2$. Since $V\geq 0$, 
the quadratic form defined by
$$
h(\ffi,\,\psi)\ =\ \bra\,\nabla\ffi,\ \nabla\psi\,\ket
\ +\ \bra\,\sqrt{V}\,\ffi,\  \sqrt{V}\,\psi\,\ket
$$
on $Q(h)=Q(-\Delta)\cap Q(V)$, is closed and positive. 
Here $Q(A)$ means the quadratic form domain of the operator $A$.
Since $D(H)\subset Q(h)$, any eigenvector of
$H$ belongs to 
$$
Q(-\Delta)\ =\ \{\,\ffi\in L^2(\R^n)\
:\ \|\,\nabla\ffi\,\|\,<\,\infty\,\}.
$$
Thus, $\nabla\ffi\in L^2$.

Next, we prove  $\ffi\in D(e^{a\bra x\ket})$, for any
$a>0$ by a Combes--Thomas argument, as presented in
Theorem XII.39 of \cite{rs4}.
We describe the details for completeness.
Let $\alpha\in\R$, and let $v$ denote $x_j$ for any
$j\in\{1,\cdots,n\}$.
We consider the unitary group 
$\ W(\alpha)\ =\ e^{i\alpha v}$ for $\alpha\in\R$,
and compute
$$
H(\alpha)\ =\ W(\alpha)\ (-\Delta + V)\ W(\alpha)^{-1}\
=\ H\ +\ i\,\alpha\,\partial_v\ +\ \alpha^2.
$$
The operator $i\partial_v$ is $H$-bounded,
with arbitrary small relative bound, since $V\geq 0$.
Thus $\{H(\alpha)\}$ extends a self-adjoint, entire
analytic family of type A, defined on $D(H)$.
We note that since $H(0)=H$ has purely discrete
spectrum, its resolvent, $R_0(\lambda)$ is compact,
for any $\lambda\in\rho(H)\equiv\C\setminus\sigma(H)$.
Hence, $R_\alpha(\lambda)=(H(\alpha)-\lambda)^{-1}$ is
compact for any $\alpha\in\R$, and hence, for all $\alpha\in\C$,
if $\lambda\in\rho(H(\alpha))$. It is
jointly analytic in $\alpha$ and $\lambda$. The eigenvalues of
$H(\alpha)$ are thus analytic in $\alpha$, except at crossing points,
where they may have algebraic singularities.
Since for $\alpha$ real, $W(\alpha)$ is unitary,
the eigenvalues are actually independent of $\alpha$,
and $\sigma(H(\alpha))=\sigma(H)$, for any $\alpha$. 

Let $P$ be the finite rank spectral projector
corresponding to an eigenvalue ${\cal E}$ of $H_{NF}$. Then, for
$\alpha\in\R$,
$P(\alpha)=W(\alpha)PW(\alpha)^{-1}$ 
is the spectral projector corresponding to the eigenvalue ${\cal E}$ of
$H(\alpha)$. By Riesz's formula and the properties of the resolvent,
$P(\alpha)$ extends to an entire analytic function that
satisfies
$$
W(\alpha_0)P(\alpha)W(\alpha_0)^{-1}=P(\alpha_0+\alpha).
$$
for any $\alpha_0\in\R$.

By O'Connor's Lemma (Sect. XIII.11 of \cite{rs4}), this yields
information about the eigenvectors.
If $\ffi=P \ffi$, the vector $\ffi^\alpha=W(\alpha)\ffi$,
defined for $\alpha\in\R$ has an
analytic extension to the whole complex plane,
and is an analytic vector for the operator $v$.
Therefore, $\ffi\in D(e^{a|v|})$, for any $a>0$. 
By taking all possible $x_j$'s for $v$, and noting that
$D(e^{a\bra x\ket})=D(e^{a(\sum_j|x_j|)})$, we see that
$\ffi\in D(e^{a\bra x\ket})$.

From this, it follows that
$\Delta\ffi\in D(e^{a\bra x\ket})$ for any $a>0$ as
well, since for any $\delta>0$,
\bea\nonumber
&&\int_{\r^n}\ e^{2a\bra x\ket}\ |\Delta \ffi(x)|^2\,dx\
\\[3mm]\nonumber 
&=&\int_{\r^n}\ e^{2a\bra x\ket}\ 
\left|\,(V(x)-{\cal E})\,\ffi(x)\,\right|^2\ dx
\\[3mm]\nonumber 
&\leq&
\|\,(V-{\cal E})^2\,e^{-\delta\bra\cdot\ket}\,\|_\infty\,\
\|\,e^{(a+\delta/2)\bra \cdot \ket}\,\ffi(\cdot)\,\|^2
\\[3mm]\nonumber 
&<&\infty.
\eea

Finally, Lemma \ref{gen} below shows that
$\nabla\ffi\in D(e^{a\bra x\ket})$.
To apply this Lemma in our situation,
we let $p(x)=e^{a\bra x\ket}$ and note that for any $a>0$,
$$
(\nabla e^{a \bra x \ket})/e^{a \bra x \ket }\ =\
a \nabla \bra x \ket\ =\ ax/\bra x \ket
$$
is uniformly bounded. \hfill\ep

\vskip 5mm
Lemma \ref{gen} requires some notation.
Letting $p(x)$ be a positive weight function,
we introduce the space 
$$
F_w^2\ =\ \left\{\,f \ : \ \| f\|_{F_w^2}^2\,=\,
\int_{\mathbb R^n}\
\left(|f(x)|^2+|\Delta f(x)|^2\right)\ p(x)\ dx
\ <\ \infty\,\right\}.
$$
We write $\|f\|_w^2\,=\,\int_{\mathbb R^n}\,|f(x)|^2\,p(x)\,dx$, 
for any $f\in L^2(\mathbb R^n, p(x)dx)$, and  
$\|f\|^2=\int_{\mathbb R^n}\,|f(x)|^2\,dx$ when
the weight is one.

\vskip 5mm
\begin{lem}\label{gen}
Let $p\in C^1$ be positive, and assume that there exists a constant 
$C<\infty$, such that 
$|(\nabla p(x))/p(x)|\,\leq\,2\,C$  
for all $x\in\mathbb R^n$. 
Then, for any $f\in F^2_w$ 
\be\label{ineq}
\|\nabla f\|_w\ \leq\ C\ \|f\|_w\ +\
\sqrt{\,\|f\|_w\ \|\Delta f\|_w\ +\ C^2\ \|f\|_w^2}.
\ee
\end{lem}

\vskip 5mm
\noindent We present the proof of this
technical lemma in Section \ref{tech}.

\vskip 5mm
We now state and prove the following Corollary
to Proposition \ref{eigv}:

\vskip 5mm
\begin{cor}\label{brr}
Assume the hypotheses of Proposition \ref{eigv}.
Let $R(\lambda)$ be the resolvent of $H=-\Delta +V$
for $\lambda\notin\sigma(H)$, and 
let $P_{{\cal E}}$ be the finite dimensional 
spectral projector of $H$ on ${\cal E}$. Let
$r({\cal E})=
\left(\,
(H-{\cal E})|_{(\mathbb I-P_{\cal E})L^2}\,
\right)^{-1}$ be the reduced 
resolvent at ${\cal E}$. Then, 
$e^{a\bra x\ket}\,R(\lambda)\,e^{-a\bra x \ket}$ and
$e^{a\bra x\ket}\,r({\cal E})\,e^{-a\bra x \ket}$
are bounded on $L^2(\R^n)$.
\end{cor}

\noindent
{\bf Proof}\quad
We use the notation of the proof of Proposition \ref{eigv}.
We know that  $R_\alpha(\lambda)$ is compact and analytic
in $\alpha \in \mathbb C$, if $\lambda\not\in \sigma(H)$.
Hence, for any $\psi_1,\,\psi_2\,\in\,C_0^\infty$, the map from
$\R\times\rho(H)$ to $\C$ given by
$$
(a, \lambda)\,\mapsto\, 
\bra\,\psi_1,\,e^{a v}\,R_0(\lambda)\,e^{-av}\,\psi_2\,\ket
$$
is uniformly bounded by $C\,\|\psi_1\|\,\|\psi_2\|$
on any given compact set of $\R\times\rho(H)$ 
for some $C$.
From this we infer that for any $a>0$,
$e^{a\bra x\ket}\,R(\lambda)\,e^{-a\bra x \ket}$ is bounded
in $L^2(\R^n)$, uniformly for $\lambda$ in compact sets of $\rho(H)$. 
Since the reduced resolvent $r({\cal E})$ can be represented as
\be\label{redres}
r({\cal E})\ =\ \frac{1}{2\pi i}\ \int_{C_{\cal E}}\
R_0(\lambda)\ \frac{1}{\lambda-{\cal E}}\ d\lambda,
\ee
where $C_{\cal E}$ is a loop in the resolvent set encircling
only ${\cal E}$, the boundedness of
$e^{a\bra x\ket}\,r({\cal E})\,e^{-a\bra x \ket}$ follows.
\hfill\ep
 
\vspace{5mm}
To show that the terms of our formal expansion all belong to $L^2$,
we use the following generalization of Proposition \ref{eigv}.
We present its proof in Section \ref{tech}.

\vskip 5mm
\begin{prop}\label{dalpha} Assume the hypotheses
of Proposition \ref{eigv} and let $\ffi$ be an $L^2$ solution of 
$(-\Delta +V-{\cal E})\,\ffi\,=\,0$.  
Then, for any $a>0$, and any multi-index $\alpha\in \mathbb N^n$, 
$\ D^\alpha\ffi\,\in\,D(e^{a\bra x\ket})$, where 
$D^\alpha\,=\,\partial_{x_1}^{\alpha_1}\,\partial_{x_2}^{\alpha_2}\,\cdots 
\,\partial_{x_n}^{\alpha_n}$. 
\end{prop}

\vskip 5mm
We now prove that our formal expansion leads to rigorous quasimodes
for the Hamiltonian $H_1(\eps)$ given by (\ref{H1}).
Theorem \ref{lead} summarizes this result for the leading order,
while Theorem \ref{asexp} handles the arbitrary order results. 

\begin{thm}\label{lead}
Let $h(\eps,\,W,\,Z)$ be defined as in Section \ref{simplest}
with $W$ shifted so that $W_0=0$.
We assume $h(\eps,\,W,\,Z)$ on ${\cal H}_{\mbox{\scriptsize el}}$
is $C^2$ in the strong resolvent sense for $(W,\,Z)$ near the origin.
We assume its non-degenerate ground state 
is given by 
\bea\nonumber
E_1(\eps,\,W,\,Z)
&=&E_0\,+\,a_1\,W^2\,+\,
\bigg(\,a_2\,\eps\,-\,a_3\,W\,\bigg)\,Z^2\,+\,
a_4\,Z^4\ +\ S(\eps,\,W,\,Z)
\\[3mm]\label{gnd}
&\equiv&
E_0\,+\,\tilde E(\eps,\,W,\,Z)\ +\ S(\eps,\,W,\,Z),
\eea
under hypothesis (\ref{hyp1}) or (\ref{hyp2}), and we
denote the corresponding normalized eigenstate by $\Phi(W,\,Z)$.
Suppose the remainder term $S$ is uniformly bounded
below by some $r>-\infty$ 
and that $|S|$ satisfies a bound of the form
\be\label{order3}
|\,S(\eps,\,W,\,Z)\,|\quad \leq\quad C\
\sum_{\alpha+\beta\ge 3}\ |\,W^\alpha\,Z^{2\beta}\,|   
\ee
for $(W,\,Z)$ in a neighborhood of the origin.
Here $C$ is independent of $\eps$, the sum is finite, and
$\alpha$ and $\beta$ are non-negative integers. 
Let $f_0(w,z)$ be a normalized non-degenerate eigenvector of $H_{NF}$,
{\it i.e.}, 
$$
(-\partial_w^2/2-\partial_z^2/2+E_{NF}(w,z))\ f_0\ =\
{\cal E}_2\ f_0,
$$ 
with
$$
E_{NF}(w,z) =\ a_1\ w^2\ +\ 
\bigg(\,a_2\ -\ a_3\,w\,\bigg)\ z^2\ +\ a_4\ z^4.
$$
Then, for small enough $\eps$, there exists an
eigenvalue ${\cal E}(\eps)$ of $H_1(\eps)$ which satisfies
$$
{\cal E}(\eps)\ =\ E_0\ +\ \eps^2\,{\cal E}_2\ +\ O(\eps^{\xi}),
$$
for some $\xi>2$ as $\eps\ra 0$.
\end{thm}

\vskip 5mm\noindent
{\bf Remarks}\quad
{\bf 1.}\quad At this level of approximation, it is not necessary to
require the eigenvector $\Phi$ to satisfy
condition (\ref{connection}) or to require $h(\eps,\,W,\,Z)$ be real 
symmetric.\\
{\bf 2.}\quad We have stated our results for the electronic ground state,
but the analogous results would be true for any non-degenerate state
that had the same type of dependence on $\eps$.

\vskip 5mm
\noindent 
{\bf Proof:}\quad
In the course of the proof, we denote all generic
non-negative constants by the same symbol $c$. 

Our candidate for the construction of a quasimode is 
\be\label{candidate}
\Psi_Q(\eps,\,W,\,Z)\quad =\quad F(W/\eps^{\delta_1})\
F(Z/\eps^{\delta_2})\ f_0(W/\eps,\,Z/\sqrt{\eps})\ \Phi(W,\,Z),
\ee 
where $F:\R\ra [0,1]$ is a smooth, even cutoff function
supported on $[-2,2]$ which is equal to $1$ on $[-1,1]$.
One should expect the introduction of these cutoffs not to
affect the expansion at any finite order because the
eigenvectors of $H_1(\eps)$ are localized near the
minimum of $E_1(\eps,\,W,\,Z)$. 
Thus, the properties of the electronic
Hamiltonian for large values $(W,\,Z)$ should not matter.
The choice  of a different cutoff for each variable
is required because these variables
have different scalings in $\eps$. We determine the precise
values of the positive exponents $\delta_1$ and $\delta_2$
in the course of the proof. We also use the notation 
\be\label{cutoff}
{\cal F}(\eps,W,Z)\quad =\quad
F(W/\eps^{\delta_1})\ F(Z/\eps^{\delta_2}).
\ee

We first estimate the norm of $\Psi_Q$.
\bea\nonumber
\|\Psi_Q\|^2&=&\int_{\r^2}\ |{\cal F}(\eps,W,Z)\
f_0(W/\eps,\,Z/\sqrt{\eps})|^2\ 
\|\Phi(W,Z)\|_{{\cal H}_{\mbox{\scriptsize el}}}^2
\\[3mm] \nonumber
&=&\int_{\r^2}\ |f_0(W/\eps,\,Z/\sqrt{\eps})|^2\ dW\,dZ
\\[3mm] \nonumber
&&\qquad\qquad\qquad\qquad-\ \int_{\r^2}\ (1-{\cal F}^2(\eps,W,Z))\
|f_0(W/\eps,\,Z/\sqrt{\eps})|^2\ dW\,dZ.
\eea
The first term of the last expression equals $\eps^{3/2}$,
by scaling, since $f_0$ is normalized.
If $\delta_1<1$ and $\delta_2<1/2$,
the negative of the second term is bounded above by
\bea\nonumber
&&\int_{|W|\geq \eps^{\delta_1} \atop |Z|\geq \eps^{\delta_2}}\
|f_0(W/\eps,\,Z/\sqrt{\eps})|^2\ dW\,dZ
\\[3mm] \nonumber
&&\quad\quad =\quad\eps^{3/2}\
\int_{|w|\geq \eps^{1-\delta_1} \atop |z|\geq \eps^{1/2-\delta_2}}\
e^{-2a(|w|+|z|)}\ e^{2a(|w|+|z|)}\ |f_0(w,\,z)|^2\ dw\,dz
\\[4mm] \nonumber
&&\quad\quad\leq\quad
\eps^{3/2}\ e^{-2a(1/\eps^{(1-\delta_1)}+1/\eps^{(1-\delta_2)})}\
\|e^{a(|\cdot|+|\cdot|)}f_0\|^2\\[3mm]\nonumber
&&\qquad=\quad O(\eps^\infty), 
\eea
since $f_0\in D(e^{a\bra (W,Z)\ket})$.
Hence,
\be\label{npq}
\|\Psi_Q\|\ =\ \eps^{3/4}\ (1+O(\eps^\infty)),\quad
\mbox{where the }O(\eps^\infty)\ \mbox{correction is non-positive.}
\ee

Next we compute
\bea\label{h-e}
&&(H_1(\eps)-(E_0+\eps^2{\cal E}_2))\ \Psi_Q(\eps,\,W,\,Z)
\\[2mm] \nonumber
&&=\quad S(\eps,\,W,\,Z)\ f_0(w,z)|_{W,Z}\ {\cal F}(\eps,W,Z)\ \Phi(W,Z)
\\[3mm] \nonumber
&&\qquad\quad-\ \left(\left(\frac{\eps^4}{2}\partial_W^2+
\frac{\eps^3}{2}\partial_Z^2\right)\
{\cal F}(\eps,W,Z)\ \Phi(W,Z)\right)\ f_0(w,z)|_{W,Z},
\\[3mm] \nonumber
&&\qquad\quad -\ \eps^{3}\ \left(\partial_wf_0(w,z)|_{W,Z}\right)\
\partial_W({\cal F}(\eps,W,Z)\Phi(W,Z))
\\[3mm] \nonumber
&&
\qquad\quad-\ \eps^{5/2}\ \left(\partial_zf_0(w,z)|_{W,Z}\right)\
\partial_Z({\cal F}(\eps,W,Z)\Phi(W,Z)),
\eea
where we have introduced the shorthand
$f_0(w,z)|_{W,Z}\,=\,f_0(W/\eps,\,Z/\sqrt{\eps})$ and used the identity
$$
\tilde E(\eps,\,W,\,Z)\ -\ \eps^2\,E_{NF}(W/\eps,\,Z/\sqrt{\eps})\
\equiv\ 0.
$$
Also
\bea\nonumber
\partial_W{\cal F}(\eps,W,Z)&=&
\frac{1}{\eps^{\delta_1}}\ F'(W/\eps^{\delta_1})\ F(Z/\eps^{\delta_2})
\\[3mm] \nonumber 
\partial_Z{\cal F}(\eps,W,Z)&=&
\frac{1}{\eps^{\delta_2}}\ F(W/\eps^{\delta_1})\ F'(Z/\eps^{\delta_2})
\eea
and, by assumption,
$\|\partial_W^\mu\partial_Z^\nu\Phi(W,Z)\|
_{{\cal H}_{\mbox{\scriptsize el}}}$
is continuous and of
order $\eps^0$ in a neigborhood of the origin, for $\mu+\nu\leq 2$.
Therefore,
\bea\nonumber
&&\sup_{\r^2}\quad
\|\,\partial_W\,({\cal F}(\eps,W,Z)\Phi(W,Z))\,\|
_{{\cal H}_{\mbox{\scriptsize{el}}}}
\quad\leq\quad\frac{c}{\eps^{\delta_1}}
\\[2mm] \label{nv}
&&\sup_{\r^2}\quad
\|\,\partial_Z\,({\cal F}(\eps,W,Z)\Phi(W,Z))\,\|
_{{\cal H}_{\mbox{\scriptsize{el}}}}
\quad\leq\quad\frac{c}{\eps^{\delta_2}}
\\[2mm] \nonumber
&&\sup_{\r^2}\quad
\|\,\partial_W^2\,({\cal F}(\eps,W,Z)\Phi(W,Z))\,\|
_{{\cal H}_{\mbox{\scriptsize{el}}}}
\quad\leq\quad\frac{c}{\eps^{2\delta_1}}
\\[2mm] \nonumber
&&\sup_{\r^2}\quad
\|\,\partial_Z^2\,({\cal F}(\eps,W,Z)\Phi(W,Z))\,\|
_{{\cal H}_{\mbox{\scriptsize{el}}}}
\quad\leq\quad\frac{c}{\eps^{2\delta_2}},
\eea
where all vectors are supported in
$\{\,(W,Z)\,:\,|W|\leq 2/\eps^{\delta_1},\,|Z|\leq 2/\eps^{\delta_2}\}$.
Each of these vectors appears in 
(\ref{h-e}), multiplied by one of the scalar functions
$f_0(w,z)|_{W,Z}$, $\ \left(\partial_wf_0(w,z)\right)|_{W,Z}$, or 
$\ \left(\partial_zf_0(w,z)\right)|_{W,Z}$.
In turn, each of these functions belongs to
$L^2(\R^2)$ by Proposition \ref{eigv}, and
each one has norm of order $\eps^{3/4}$ because of scaling, {\it e.g.},
$$
\left(\,\int_{\r^2}\ |\,(\partial_wf_0)(W/\eps,\,Z/\sqrt{\eps})\,|^2\
dW\,dZ\,\right)^{1/2}\quad =\quad
\eps^{3/4}\ \|\,\partial_wf_0\,\|_{L^2(\r^2)}.
$$
Therefore, the norms of the last three vectors in (\ref{h-e}) are of
order $\eps^{3/4}$ times the corresponding power of $\eps$ stemming
from (\ref{nv}).

We now estimate the norm of the term that arises from the error term
$S$. From our hypothesis on the behavior of $S$, we have 
\bea\nonumber
\|\,S\ {\cal  F}\ f_0\ \Phi\,\|^2&=&
\int_{|W|\leq 2/\eps^{\delta_1} \atop |Z|\leq 2/\eps^{\delta_2}}\
|f_0(W/\eps,\,Z/\sqrt{\eps})\ S(W,Z)|^2\ dW\,dZ
\\[3mm]\nonumber
&\leq& c\ \sum_{\alpha+\beta\ge 3}\
\int_{|W|\leq 2/\eps^{\delta_1} \atop |Z|\leq 2/\eps^{\delta_2}}\
|f_0(W/\eps,\,Z/\sqrt{\eps})|^2\
\left|\,W^\alpha Z^{2\beta}\,\right|^2\ dW\,dZ
\\[3mm]\nonumber
&\leq&c\ \sum_{\alpha+\beta\ge 3}\
\eps^{2(\alpha\delta_1+2\beta\delta_2)}\ \int_{\r^2}\
|f_0(W/\eps,\,Z/\sqrt{\eps})|^2\
dW\,dZ
\\[3mm]\nonumber
&=&c\ \sum_{\alpha+\beta\ge 3}\
\eps^{2(\alpha\delta_1+2\beta\delta_2)}\ \eps^{3/2},
\eea
where the sums are finite.

Collecting these estimates and inserting the allowed values
of $\alpha$ and $\beta$, we obtain
\bea\nonumber
&&\|\,(H_1(\eps)-(E_0+\eps^2{\cal E}_2))\ \Psi_Q\,\|
\\[3mm] \nonumber
&&\leq\quad c\ \eps^{3/4}\ \left(\,\eps^{3\delta_1}+
\eps^{2(\delta_1+\delta_2)}+\eps^{\delta_1+4\delta_2}+
\eps^{6\delta_2}+\eps^{4-2\delta_1}+\eps^{3-2\delta_2}
+\eps^{3-\delta_1}+\eps^{5/2-\delta_2}\,\right).
\eea
We further note that $\delta_1<1$ and $\delta_2<1/2$ imply 
$\eps^{4-2\delta_1}\ll \eps^{3-\delta_1}$ and
$\eps^{3-2\delta_2}\ll \eps^{5/2-\delta_2}$. This, together with
(\ref{npq}), shows that for small enough $\eps$,
$$
\frac{\|\,(H_1(\eps)-(E_0+\eps^2{\cal E}_2))\ \Psi_Q\,\|}
{\|\,\Psi_Q\,\|}\quad\leq\quad 
c\ \left(\eps^{3\delta_1}+\eps^{2(\delta_1+\delta_2)}+
\eps^{\delta_1+4\delta_2}+ \eps^{6\delta_2}
+\eps^{3-\delta_1}+\eps^{5/2-\delta_2}\right)
$$

We still must show that all terms in the parenthesis above
can be made asymptotically smaller than $\eps^2$. 
This can be done if there exist choices of 
$\delta_1$ and $\delta_2$ such that all exponents in the
parenthesis above are strictly larger than 2.
The inequalities to be satisfied are 
\bea\nonumber
0<\delta_1<1,\quad\delta_1>2/3,\quad\delta_1+\delta_2>1,\quad
0<\delta_2<1/2,\quad\delta_2>1/3,\quad\delta_1+4\delta_2>2.
\eea
Satisfying these is equivalent to satisfying 
$$
\left\{\matrix{2/3<\delta_1<1\vspace{2mm}\cr 1/3 <\delta_2 <1/2}\right.
$$
which defines the set of allowed values. The best value,
$$
\xi\quad =\quad
\max_{0<\delta_1<1\atop 0<\delta_2<1/2}\min\
\left\{\,3\delta_1,\ 2(\delta_1+\delta_2),\ \delta_1+4\delta_2,\
6\delta_2,\ 3-\delta_1,\ 5/2-\delta_2\,\right\}\quad>\quad 0,
$$
is obtained by straighforward optimization and is given by $\xi=15/7$,
obtained for $5/7<\delta_1<6/7$ and $\delta_2=5/14$.
With such a choice, there exists an eigenvalue ${\cal E}(\eps)$
of $H_1(\eps)$ that satisfies
$$
{\cal E}(\eps)\ =\ E_0\ +\ \eps^2\,{\cal E}_2(\eps)\ +\ O(\eps^{\xi}),
$$
with $\xi=2+1/7$.
\hfill\ep

\vskip 5mm
We now turn to the construction of a complete asymptotic expansion
for the energy level ${\cal E}(\eps)$ of $H_1(\eps)$, as $\eps\ra 0$.

\begin{thm}\label{asexp}
Assume the hypotheses of Theorem \ref{lead} with the additional
condition that $h(\eps,\,W,\,Z)$ on ${\cal H}_{el}$ is
$C^\infty$ in the strong resolvent sense in the
variables $(\eps,\,W,\,Z)$. 
Then the energy level ${\cal E}(\eps)$ of $H_1(\eps)$
admits a complete asymptotic expansion in powers of $\eps^{1/2}$.
The same conclusion is true for the 
corresponding quasimode eigenvector.
\end{thm}

\noindent
{\bf Proof}\quad
Our candidate for the quasimode is again the formal expansion 
(\ref{forsol}) truncated at order $\eps^{N/2}$ 
and multiplied by the cutoff function (\ref{cutoff}), {\it i.e.}, 
$$
\Psi_Q(\eps,\,W,\,Z)\quad =\quad {\cal F}(\eps,\,W,\,Z)\quad
\sum_{j=0}^N\ \eps^{j/2}\ \psi_{j/2}(W,\,Z,\,W/\eps,\,Z/\sqrt{\eps}).
$$

We shall determine $\psi_{j/2}$ and $T_{j/2}$ in (\ref{H2}) explicitly, 
but first we introduce
some notation for certain Taylor series. Expanding in powers of
$\eps^{1/2}$, we write 
\bea\nonumber\hspace{-13mm}
&&T_{j/2}(W,\,Z)\ -\ T_{j/2}(\eps w,\,\eps^{1/2}z)
\\[2mm] \nonumber\hspace{-13mm}
&=&T_{j/2}(W,Z)-T_{j/2}(0,0)
\,-\,\eps^{1/2}\partial_Z T_{j/2}(0,0)\,+\,
\eps\left(\partial_W T_{j/2}(0,0)w+\partial_{Z^2}T_{j/2}(0,0)w^2/2\right)
+\cdots
\\[2mm] \nonumber\hspace{-13mm} 
&\equiv&T_{j/2}(W,\,Z)\ -\ T_{j/2}(0,\,0)\ +\
\sum_{k=1}^\infty\ \tau^{(k/2)}_{j/2}(w,\,z)\ \eps^{k/2}.
\eea
Next, our hypotheses imply that the function
$S(\eps, W, Z)$ in (\ref{gnd})
$$
E_1(\eps, W, Z)\quad =\quad
E_0\ +\ \tilde E(\eps,\,W,\,,Z)\ +\ S(\eps,\,W,\,Z)
$$ 
is $C^\infty$ in $(\eps,W,Z)$. Using (\ref{order3}), we write 
$$
E_1(\eps,\,\eps w,\,\eps^{1/2}z)\quad =\quad
E_0\ +\ \eps^2\,E_{NF}(w,\,z)\ +\
\sum_{m\geq 6}^\infty\ \eps^{m/2}\ S_{m/2}(w,\,z).
$$

\vskip 3mm \noindent
{\bf Note}\quad Because we have assumed $E_1(\eps,\,W,\,Z)$
is even in $Z$, $\ S_{m/2}(w,\,z)=0$ when $m$ is odd,
but the notation is somewhat simpler if we include these terms.

\vskip 3mm
We use this notation and substitute the formal series
(\ref{forsol}) and (\ref{expe}) into the eigenvalue equation
(\ref{MSequation}), with $H_2$ given by (\ref{H2}).
For orders $n/2$ with $n\le 4$, we find exactly what we obtained in
Section \ref{simplest}.
When $n\ge 5$, we have to solve
\bea\label{fullexp}\hspace{-10mm}
&&[h(\eps,\,W,\,Z)-E_1(\eps,\,W,\,Z)]\ \psi_{n/2}
\\[3mm] \nonumber\hspace{-10mm}
&+&E_{NF}(w,z)\psi_{(n-4)/2}+S_{6/2}(w,z)\psi_{(n-6)/2}
+S_{7/2}(w,z)\psi_{(n-7)/2}
+\cdots+S_{n/2}(w,z)\psi_0
\\[3mm] \nonumber\hspace{-10mm}
&+&(T_{\frac{n}{2}}(W,Z)-T_{\frac{n}{2}}(0,0))\psi_0-
\tau_{\frac{n-1}{2}}^{(\frac12)}(w,z)\psi_0
-\tau_{\frac{n-2}{2}}^{(\frac22)}(w,z)\psi_0-
\cdots -\tau_{\frac62}^{(\frac{n-6}{2})}(w,z)\psi_0
\\[3mm] \nonumber\hspace{-10mm}
&+&(T_{\frac{n-1}{2}}(W,Z)-T_{\frac{n-1}{2}}(0,0))\psi_{1/2}-
\tau_{(n-2)/2}^{(1/2)}(w,z)\psi_{1/2}
-\cdots -\tau_{\frac62}^{(\frac{n-7}{2})}(w,z)\psi_{1/2}
\\[3mm] \nonumber\hspace{-10mm}
&& \qquad  \qquad \qquad \qquad \vdots
\\[3mm] \nonumber\hspace{-10mm}
&+&(T_{\frac72}(W,Z)-T_{\frac72}(0,0))\psi_{\frac{n-7}{2}}-
\tau_{\frac62}^{(\frac12)}(w,z)\psi_{(n-7)/2}
\\[3mm] \nonumber\hspace{-10mm}
&+&(T_{\frac62}(W,Z)-T_{\frac62}(0,0))\psi_{(n-6)/2}
\\[3mm] \nonumber\hspace{-10mm}
&-&\frac12\Delta_{w,z}\psi_{(n-4)/2}-\partial_{Z,z}^2\psi_{(n-5)/2}-
\partial_{W,w}^2\psi_{(n-6)/2}-\frac12\partial_{Z,Z}^2\psi_{(n-6)/2}-
\frac12\partial_{W,W}^2\psi_{(n-8)/2}
\\[3mm] \nonumber\hspace{-10mm}
&=&{\cal E}_2\ \psi_{(n-4)/2}\ +\ {\cal E}_{5/2}\ \psi_{(n-5)/2}
\ +\ \cdots\ +\ {\cal E}_{n/2}\psi_0,
\eea 
with the understanding that the quantities $S$,
$T$ and $\tau$ that appear with indices lower than
those allowed in their definitions are equal to zero.

We solve (\ref{fullexp} by induction on $n$. We assume that
$$\left\{\begin{array}{lcl}
{\cal E}_{j/2},\quad\psi^\perp_{j/2}(W,Z,w,z),
\quad T_{j/2}(W,Z)\quad&\mbox{for} 
&j\leq n-1,\qquad\ \mbox{and}\vspace{3mm}\cr
f_{j/2}(W,Z,w,z)&\mbox{for}&j\leq n-5
\end{array}\right.
$$
have already been determined, with $f_{j/2}^\parallel(W,Z,w,z)=0$,
for $j\geq 1$.

We project (\ref{fullexp}) into the $\Phi(W,\,Z)$ direction
and the orthogonal direction in the electronic Hilbert space
to obtain two equations that must each be solved.

First, we take the scalar product of (\ref{fullexp})
with $\Phi(W,Z)$ in the electronic Hilbert 
space to obtain
\bea\nonumber\hspace{-10mm}
&&\hspace{-7mm}E_{NF}(w,z)f_{(n-4)/2}\,+\,S_{6/2}(w,z)f_{(n-6)/2}
\,+\,S_{7/2}(w,z)f_{(n-7)/2}\,+\,\cdots\,+\,S_{n/2}(w,z)f_0
\\[3mm] \nonumber\hspace{-10mm}
&+&(T_{\frac{n}{2}}(W,\,Z)-T_{\frac{n}{2}}(0,\,0))f_0\ -\
\tau_{\frac{n-1}{2}}^{(\frac12)}(w,z)f_0\ -\
\tau_{\frac{n-2}{2}}^{(\frac22)}(w,z)f_0\ -\
\cdots\ -\ \tau_{\frac62}^{(\frac{n-6}{2})}(w,z)f_0
\\[3mm] \nonumber\hspace{-10mm}
&+&(T_{\frac{n-1}{2}}(W,\,Z)-T_{\frac{n-1}{2}}(0,\,0))f_{1/2}\ -\
\tau_{(n-2)/2}^{(1/2)}(w,z)f_{1/2}\ -\
\cdots\ -\ \tau_{\frac62}^{(\frac{n-7}{2})}(w,z)f_{1/2}
\\[3mm] \nonumber\hspace{-10mm}
&& \qquad  \qquad \qquad \qquad \vdots
\\[3mm] \nonumber\hspace{-10mm}
&+&(T_{\frac72}(W,\,Z)-T_{\frac72}(0,\,0))\ f_{\frac{n-7}{2}}
\ -\ \tau_{\frac62}^{(\frac12)}(w,z)\ f_{(n-7)/2}
\\[3mm] \nonumber\hspace{-10mm}
&+&(T_{\frac62}(W,\,Z)-T_{\frac62}(0,\,0))\ f_{(n-6)/2}
\\[3mm] \nonumber\hspace{-10mm}
&-&\frac12\ \Delta_{w,z}f_{(n-4)/2}\ -\
\bra\Phi(W,\,Z),\,\partial_{Z,z}^2\psi_{(n-5)/2}\ket_{{\cal H}_{el}}\ -\
\bra\Phi(W,\,Z),\,\partial_{W,w}^2\psi_{(n-6)/2}\ket_{{\cal H}_{el}}
\\[3mm] \nonumber\hspace{-10mm}
&-&\frac12\ \bra\Phi(W,\,Z),\,
\partial_{Z,Z}^2\psi_{(n-6)/2}\ket_{{\cal H}_{el}}
\ -\ \frac12\ \bra\Phi(W,\,Z),\,
\partial_{W,W}^2\psi_{(n-8)/2}\ket_{{\cal H}_{el}}
\\[3mm]\hspace{-10mm}
&=&{\cal E}_2\ f_{(n-4)/2}\ +\ {\cal E}_{5/2}\ f_{(n-5)/2}\ +\
\cdots\ +\ {\cal E}_{n/2}\ f_0.
\label{paralexp}
\eea

We further project (\ref{paralexp}) into the $f_0$ direction
and the orthogonal direction in $L^2(\R^2)$
to obtain two equations that must each be solved.

We take the scalar product of (\ref{paralexp}) with $f_0$
in $L^2(\mathbb R^2)$.
Using $f_{j/2}^\parallel=0$ for $j\geq 1$ and  
$(-\frac12\Delta_{w,z}+E_{NF}(w,z))\,f_0\ =\
{\cal E}_{2}\,f_0$, we obtain
\bea\label{paralparalexp}\hspace{-10mm}
{\cal E}_{n/2}
&=&T_{\frac{n}{2}}(W,\,Z)-T_{\frac{n}{2}}(0,\,0)
\\[3mm] \nonumber\hspace{-10mm}
&&+\quad\sum_{j=6}^n\
\bra\,f_0,\,S_{j/2}\ f_{(j-6)/2}\,\ket_{L^2(\mathbb R^2)}\ -\
\sum_{k=0}^{n-7}\ \sum_{j=1}^{n-6-k}\
\bra\,f_0,\,\tau_{j/2}^{(n-(j+k))/2)}f_{k/2}\,\ket_{L^2(\mathbb R^2)}
\\[3mm] \nonumber\hspace{-10mm}
&&-\quad\left\bra\,f_0,\,
\bigg\{\bra\,\Phi(W,\,Z),\,
\partial_{Z,z}^2\psi_{(n-5)/2}\ket_{{\cal H}_{el}}+
\bra \Phi(W,Z),\,
\partial_{W,w}^2\psi_{(n-6)/2}\ket_{{\cal H}_{el}}\right.
\\[3mm] \nonumber\hspace{-10mm}
&&\qquad\qquad+\ \frac12\,\bra \Phi(W,Z),\,
\partial_{Z,Z}^2\psi_{(n-6)/2}\ket_{{\cal H}_{el}}+
\frac12\,\left.\bra\Phi(W,Z),\,
\partial_{W,W}^2\psi_{(n-8)/2}\ket_{{\cal H}_{el}}\bigg\}
\right\rangle_{L^2(\mathbb R^2)}.
\eea
We can solve this equation for ${\cal E}_{n/2}$ if the
right hand side is  independent of $(W,\,Z)$.
This will be true, if we choose
\bea\nonumber\hspace{-1cm}
T_{\frac{n}{2}}(W,\,Z)&=&-\
\sum_{j=6}^n\ \bra\,f_0,\,S_{j/2}\
f_{(j-6)/2}\,\ket_{L^2(\mathbb R^2)}
\ +\ \sum_{k=0}^{n-7}\ \sum_{j=1}^{n-6-k}
\bra\,f_0,\,\tau_{j/2}^{(n-(j+k))/2)}\
f_{k/2}\,\ket_{L^2(\mathbb R^2)}
\\[3mm] \nonumber\hspace{-1cm}
&+&\left\bra f_0,\ \bigg\{
\bra\Phi(W,Z),\,
\partial_{Z,z}^2\psi_{(n-5)/2}\ket_{{\cal H}_{el}}+
\bra\Phi(W,Z),\,
\partial_{W,w}^2\psi_{(n-6)/2}\ket_{{\cal H}_{el}}\right.
\\[3mm] \nonumber\hspace{-1cm}
&&\qquad\
+\ \frac12\bra\Phi(W,Z),\,
\partial_{Z,Z}^2\psi_{(n-6)/2}\ket_{{\cal H}_{el}}+
\frac12\left.\bra\Phi(W,Z),\,
\partial_{W,W}^2\psi_{(n-8)/2}\ket_{{\cal H}_{el}}
\bigg\}\right\ket_{L^2(\mathbb R^2)}
\eea
We then are forced to take
$$
{\cal E}_{n/2}\ =\ -\ T_{\frac{n}{2}}(0,\,0).
$$
The first non-zero $T_{j/2}(W,Z)$ is
\be\label{T6/2}
T_{6/2}(W,\,Z)\ =\ \frac12\
\bra\,\Phi(W,\,Z),\,\partial_Z^2\,\Phi(W,\,Z)\,\ket_{{\cal H}_{el}}
\ -\ \bra\,f_0,\,S_3\ f_0\,\ket_{L^2(\r ^2)}.
\ee
So,
$$
{\cal E}_3\ =\ \bra\,f_0,\,S_3\ f_0\,\ket_{L^2(\r ^2)}\ -\
\frac12\ \bra\,\Phi(0,\,0),\,
(\partial_Z^2\,\Phi)(0,\,0)\,\ket_{{\cal H}_{el}}.
$$

We next equate the components on the two sides of
(\ref{paralexp}) that are orthogonal to
$f_0$ in $L^2(\R^2)$. The resulting equation can be solved by
applying the
reduced resolvent $r_{NF}({\cal E}_2)$, which is the
inverse of the restriction of
$(-\frac12\Delta_{w,z}+E_{NF}-{\cal E}_2)$ to the subspace
orthogonal to $f_0$. We thus obtain
\bea\label{fperp}\nonumber\hspace{-1cm}
f_{(n-4)/2}&=& r_{NF}({\cal E}_2)\ \left[\ \sum_{j=1}^{n-5}\
{\cal E}_{(n-j)/2}\ f_{j/2}^\perp\
-\ \sum_{j=6}^n\ (S_{j/2}(w,z)\ f_{(j-6)/2})^\perp \right.
 \\ \nonumber\hspace{-1cm}
&+&\sum_{k=0}^{n-7}\sum_{j=1}^{n-6-k}
(\tau_{j/2}^{(n-(j+k))/2)}(w,z)f_{k/2})^\perp\,+\,
\sum_{j=1}^{n-6}(T_{(n-j)/2}(0,0)-T_{(n-j)/2}(W,Z))f_{j/2}^\perp 
\\ \nonumber\hspace{-1cm}
&+&
\bra\,\Phi(W,\,Z),\,\partial^2_{Z,z}\,\Psi_{(n-5)/2}\,
\ket_{{\cal H}_{el}}^\perp\
+\ \bra\,\Phi(W,\,Z),\,\partial^2_{W,w}\,\Psi_{(n-6)/2}\,
\ket_{{\cal H}_{el}}^\perp 
\\ \nonumber\hspace{-1cm}
&+&\left.
\bra\,\Phi(W,\,Z),\,\partial^2_{Z,Z}\,\Psi_{(n-6)/2}\,
\ket_{{\cal H}_{el}}^\perp\
+\ \bra\,\Phi(W,\,Z),\,\partial^2_{W,W}\,\Psi_{(n-8)/2}\,
\ket_{{\cal H}_{el}}^\perp
\phantom{\sum_{j=1}^{n-5}}\hspace{-.5cm}\right], 
\eea
This solution has $f_{(n-4)/2}\,=\,f_{(n-4)/2}^\perp$
orthogonal to $f_0$, as claimed in Section \ref{simplest}.
The first non-trivial $f_{j/2}$, for $j\geq 1$ is 
\be\label{fperp2/2}
f_{1}^\perp(W,\,Z,\,w,\,z)\ =\ -\ r_{NF}({\cal E}_2)\
(S_3(w ,z)\,f_0(w, z))^\perp.
\ee

Next, we equate the components of (\ref{fullexp})
that are orthogonal to $\Phi(W,\,Z)$ in ${\cal H}_{el}$.
We solve the resulting equation for
$\psi_{n/2}^\perp$by applying the reduced resolvent
$r(W,\,Z)$ of $h(\eps,\,W,\,Z)$ at $E_1(\eps, W, Z)$. This yields
\bea\label{psiperp}\hspace{-1cm}
\psi_{n/2}^\perp &=& r(W,Z)\left[\
\sum_{j=1}^{n-5}\ {\cal E}_{(n-j)/2}\ \psi_{j/2}^\perp
\ -\ \sum_{j=6}^n\ S_{j/2}(w,z)\ \psi_{(j-6)/2}^\perp\right.
\\[3mm] \nonumber\hspace{-1cm}
&&\qquad +\ \sum_{k=0}^{n-7}\sum_{j=1}^{n-6-k}
\tau_{j/2}^{(n-(j+k))/2)}(w,z)\psi_{k/2}^\perp + \sum_{j=1}^{n-6}
(T_{(n-j)/2}(0,0)-T_{(n-j)/2}(W,Z))\psi_{j/2}^\perp 
\\[3mm] \nonumber\hspace{-1cm}
&&\qquad+\
(\partial^2_{Z,z}\,\psi_{(n-5)/2})^\perp\ +\
((\partial^2_{W,w} +\partial^2_{Z,Z})\ \psi_{(n-6)/2})^\perp 
\\[3mm] \nonumber\hspace{-1cm}
&&\qquad+\ \left.
(\partial^2_{W,W}\ \psi_{(n-8)/2})^\perp\ -\ 
(\frac12\Delta_{w,z}+E_{NF}(w,z)-{\cal E}_{2})\
\psi_{(n-4)/2}^\perp\phantom{\sum_{j=1}^{n-5}}\hspace{-.5cm}\right].
\eea
The first non-zero component $\psi^\perp_{j/2}$ with $j\geq 0$, is 
\be\label{psiperp5/2}
\psi_{5/2}^\perp (W,\,Z,\,w,\,z)\quad =\quad
(\partial_z f_0)(w,z)\ r(W,\,Z)\ (\partial_Z\Phi)(W,\,Z).
\ee

\vskip 5mm
Finally, Proposition \ref{doghouse} below shows that each
$\psi_{j/2}$ in this expansion belongs to 
$D(e^{a(|W|/\eps+|Z|/\sqrt{\eps})})$.
As a result, whenever a derivative 
acts on the cutoff, it yields a contribution whose $L^2$ norm
is exponentially small. This way, we can neglect such terms.
For example 
$$
(\partial_W {\cal F}(\eps,\,W,\,Z))\,
\psi_{j/2}(W,\,Z,\,W/\eps,\,Z/\sqrt{\eps})\,=\,
{\eps^{-\delta_1}}\,F'(W/\eps^{\delta_1})\,F(Z/\eps^{\delta_2})\,
\psi_{j/2}(W,\,Z,\,W/\eps,\,Z/\sqrt{\eps}).
$$
The square of the  $L^2$ norm of this term
is bounded by a constant times
$$
\eps^{-2\delta_1}\int_{|W|/\eps^{\delta_1}\geq 1,
|Z|/\eps^{\delta_2}\leq 2}|
\psi_{j/2}(W,Z,W/\eps, Z/\sqrt{\eps})|^2
e^{2a (|W|/\eps+|Z|/\sqrt{\eps})}
e^{-2a (1/\eps^{1-\delta_1})}dWdZ\
=\ O(\eps^{\infty}).
$$

Thus, we have constructed the non-zero 
quasimode (\ref{candidate}) that satisfies the eigenvalue equation 
up to an arbitrary high power of $\eps^{1/2}$.\hfill\ep

\vskip 5mm
The proof of Proposition \ref{doghouse} relies on the following lemma.

\vskip 5mm
\begin{lem}\label{lem}Let $V$ be a polynomial that is bounded
below, such that the spectrum of $H=-\frac 12\Delta+V$ purely discrete.
Let $\ffi\in C^\infty(\R^n)$ satisfy
$D^\alpha \ffi\in D(e^{a\bra x\ket})$,
for all $\alpha\in\N^n$ and any $a>0$.
If $R(\lambda)$ denotes the resolvent of $H$,
then $D^\alpha R(\lambda)\ffi \in D(e^{a\bra x\ket })$
for all $\alpha\in\N^n$ and all $\lambda$ in $\rho(H)$.
The same is true for $D^\alpha r({\cal E})\ffi$, where 
$r({\cal E})$ is the reduced resolvent at ${\cal E}$.
\end{lem}

\noindent
{\bf Proof}\quad
We first note that elliptic regularity implies that the resolvent
$R(\lambda) $ maps $C^\infty$ functions to $C^\infty$ functions.
Next, applied to smooth functions in $L^2$, we have the identity
$$
[\,\partial_{x_j},\,R(\lambda)\,]\ =\
R(\lambda)\ (\partial_{x_j}V)\ R(\lambda).
$$
We claim that the operators on the two sides of this equation have
bounded extensions to all of $L^2$. 
To see this, note that $D^\beta V$ is relatively bounded 
with respect to $V$ for any $\beta\in\N^n$, because $V$ is a polynomial.
Furthermore, since $H\,\ge\,V$,
we see that $D^\beta V$ is relatively bounded with respect to $H$,
which implies the claim.
Hence, for $\ffi$ as in the lemma, we have
\be\label{cdr}
\partial_{x_j}\ R(\lambda)\ \psi\ =\ R(\lambda)\ \partial_{x_j}\ \ffi
\ +\ R(\lambda)\ (\partial_{x_j}V)\ R(\lambda)\ \ffi.
\ee 
The first term on the right hand side of this equation
belongs to $D(e^{a\bra x\ket })$ since
$R(\lambda)$ maps exponentially decaying functions to exponentially
decaying functions (see Corollary \ref{brr}).
The same is true for the second 
term, with a possible arbitrarily small loss
on the exponential decay rate, due to 
the polynomial growth of $\partial_{x_j}V$.
This provides the starting point for an induction
on the order of the derivative
that appears in the conclusion of the lemma.

We now assume that for some $\alpha\in\N^n$, $D^\alpha R(z)\ffi$
is a linear combination of smooth functions of the form 
$R(z)(D^{\gamma_1}V)R(z)\cdots
R(z)(D^{\gamma_{j-1}}V)R(z)D^{\gamma_j}\ffi$
all of which belong to $D(e^{a\bra x\ket})$, for any $a>0$.
We assume every $\gamma$ that occurs here has $|\gamma|\leq|\alpha|$.
Then $\partial_{x_j}D^\alpha R(z)\ffi$ is a linear 
combination of elements of the form
$$
\partial_{x_j}(R(z)(D^{\gamma_1}V)R(z)\cdots
R(z)(D^{\gamma_{j-1}}V)R(z)D^{\gamma_j}\ffi)
$$
Applying (\ref{cdr}) successively,
we see that the structure is preserved. 
Since all $D^{\gamma_k}V$ are polynomial,
Corollary \ref{brr} implies the result.

The statement for the reduced resolvent follows 
from the representation (\ref{redres}).\hfill\ep

\vskip 5mm
\begin{prop}\label{doghouse}
Assume the hyptheses of Theorem \ref{asexp}.
Let $\psi_{j/2}(W,\,Z,\,w,\,z)$ be determined
by the construction above,
where $(W,\,Z)$ belongs to a closed neighborhood $\Omega$ 
of the origin and $(w,\,z)\in\R^2$.
Then $\psi_{j/2}$ is $C^\infty$, and the function
$G(w,\,z)\,=\,\sup_{(W,Z)\in\Omega}\ |\psi_{j/2}(W, Z, w, z)|$
belongs to $D(e^{a\bra(w,z)\ket})$.
\end{prop}

\noindent
{\bf Proof}\quad The hypothesis on the Hamiltonian
and the properties of the
normal form $H_{NF}$ proven above imply that $\Phi(W,\,Z)$ and 
$r(W,\,Z)$ are smooth, and that $r_{NF}({\cal E}_2)$ maps smooth 
functions to smooth functions. We also know that the non-degenerate 
eigenstate $f_0$ is smooth and belongs to $D(e^{a\bra(w,z)\ket})$.
The smoothness of  $\psi_{j/2}(W,\,Z,\,w,\,z)$ follows trivially
in $\Omega\times \R^2$.

Concerning the exponential decay, 
we observe that the $(w,\,z)$ dependence of $\psi_{j/2}$ stems from
the successive actions of derivatives, reduced resolvents, and
multiplications  by polynomials in $(w,\,z)$,
acting on the eigenstate $f_0$.
Lemma \ref{lem} applied in conjunction with
Proposition \ref{dalpha} shows that
the exponential  decay properties are preserved
under such operations.\hfill\ep


\vskip 5mm
\section{Technicalities}\label{tech}
\setcounter{equation}{0}

In this section, we present the proofs of Lemma \ref{gen} and
Proposition \ref{dalpha}.

\vskip 5mm \noindent
{\bf Proof of Lemma \ref{gen}}\quad 
We first note that the hypothesis on $p$ implies $p(x)>0$ for any
$x\in\mathbb R^n$, and that
\be\label{gradp}
e^{-2C|x-y|}\leq p(x)/p(y) \leq e^{2C|x-y|}.
\ee
Let $B_{R}\in\mathbb R^n$ be a ball of radius $R>0$. We first 
show that $f\in L^2(B_{R+1})$ and $\Delta f \in L^2(B_{R+1})$ 
imply $f\in H^2(B_R)$, where
$$
H^2(B_R)=\{\,f\in L^2(B_R),\ \nabla f \in L^2(B_R),\
\mbox{and}\ \Delta f\in L^2(B_R)\,\}.
$$
We denote the usual $H^2(B_R)$ norm by $\|\cdot\|_{H^2(B_R)}$.
We now show the existence of a constant $K(R)>0$, which depends only
on $R$, such that 
\be\label{rest}
\int_{B_R}\ |\nabla f|^2\ \leq\ K(R)\ \int_{B_{R+1}}\
\left( |\Delta f|^2+ | f|^2  \right).
\ee
{\bf Note}\quad This estimate does not hold in general if the
balls over which one integrates have the same radius.

We set $g=\Delta f$ on $B_{R+1}$ and $g(x)=0$ if $|x|>R+1$.  
We can then decompose $f=f_1+f_2$ with $f_1$ and $f_2$ solutions to
$$
\left\{\matrix{\Delta f_1=g,&&
\quad f_1|_{\partial B_{R+3}}=0\vspace{3mm}\cr
\Delta f_2=0&&\quad |x|\leq R+1.}\right. 
$$
Thus, $f_1\in H^2(B_{R+3})$, and there exists a constant $c_1(R)$,
which depends only on $R$, such that 
\be\label{fdlf}
\| f_1 \|_{H^2(B_{R+3})}\ \leq\ c_1(R)\ \|\Delta f_1\|_{L^2(B_{R+3})}
\ =\ c_1(R)\ \|\Delta f\|_{L^2(B_{R+1})},
\ee 
so that
$$
\|\nabla f_1\|_{L^2(B_{R+1})}\ \leq\
\|f_1\|_{H^2(B_{R+3})}\ \leq\ c_1(R)\ \|\Delta f\|_{L^2(B_{R+1})}.
$$
By the mean value property for harmonic functions,
$f_2$ also satisfies estimate (\ref{rest}), for some constant 
$K_2(R)$ with $\Delta f_2=0$
(see {\it e.g.}, Chapter 8 of \cite{abr}).
Combining these arguments, we see that for $c_2(R)=c_1(R)+K_2(R)$,
$$
\int_{B_R}|\nabla (f_1+f_2)|^2\ \leq\
c_2(R)\int_{B_{R+1}}(|\Delta f_1|^2+|f_2|^2)\ \leq\
c_2(R)\int_{B_{R+1}}(|\Delta f|^2+2(|f|^2+|f_1|^2)).
$$
But $\int_{B_{R+1}}\ |f_1|^2\ \leq\ \|f_1\|^2_{H^2(B_{R+3})}$, so
(\ref{fdlf}) implies that (\ref{rest}) holds for some constant $K(R)$.
 
Because of (\ref{gradp}), we can insert the weight $p$ into 
this estimate to establish the existence of another constant
$\tilde K(R)$, which depends only on $R$, such that
\be\label{prest}
\int_{B_R}\ p\ |\nabla f|^2\ \leq\ \tilde K(R)\
\int_{B_{R+1}}\ p\ (|\Delta f|^2+ |f|^2).
\ee
In other words, $p^{1/2}\,\nabla f\in L^2_{loc}$ if $f\in F_w^2$.

A second step consists in showing that $p^{1/2}\,\nabla f$ is in
$L^2(\mathbb R^n)$ and satisfies (\ref{ineq}).
Let $\chi_R\in C^\infty(\mathbb R^n)$ be a truncation function 
such that $0\leq \chi_R \leq 1$, 
with $\chi_R(x)=1$ if $|x|\leq R$, and $\chi_R(x)=0$ if $|x|\geq R+1$.
We can take $\chi_R$ so that $\|\nabla \chi_R\|_\infty$ is independent
of $R$. Let $f\in F_w^2$, and set $f_R=\chi_R f$.
Since $\nabla f_R=\chi_R\nabla f+f\nabla\chi_R$, we see that
$\|p^{1/2}\,\nabla f_R\|_{L^2(B_{R})}=\|p^{1/2}\,
\nabla f\|_{L^2(B_{R})}$, and 
$$
\lim_{R\ra \infty}\ \|p^{1/2}\,(f_R-f)\|_{L^2(\mathbb R^n)}\ra 0,
$$ 
by Lebesgue dominated convergence. By the same argument with
$\Delta f_R=\chi_R\Delta f+f\Delta\chi_R+2\nabla\chi_R\nabla f$,
$$
\lim_{R\ra \infty}\ \|p^{1/2}\,
(\Delta f\,-\,(\chi_R \Delta f + f\Delta\chi_R ))
\|_{L^2(\mathbb R^n)}\ =\ 0.
$$
We have the estimate
$\|p^{1/2}\,\nabla \chi_R\,\nabla f\|^2_{L^2(B_{R+1})}\ \leq\
c_2\ \int_{B_{R+1}\setminus B_{R}}\ p^{1/2}\ |\nabla f|^2$, 
for some constant $c_2$, independent of $R$. 
We can  cover the set $B_{R+1}\setminus B_{R}$
by a finite set of balls 
$\{B_1(j)\}_{j=1,\cdots, N(R)}$, of radius $1$,
centered at points $x_j$ such that $|x_j| = R+1/2$. 
In each of these balls $B_1(j)$, we can apply (\ref{prest})
(with a constant $\tilde K_1$, independent of $R$),
to see that
$$
\int_{B_{R+1}\setminus B_{R} }\ p\ |\nabla f|^2\ \leq\
c_2\ \tilde K_1\ \sum_{j=1}^{N(R)}\
\int_{B_2(j)}\ p\ (|\Delta f|^2+ | f|^2),
$$
where $B_2(j)$ has radius 2 instead of 1.
Using $\cup_{j=1}^{N(R)}B_2(j)\subset B_{R+3}\setminus B_{R-3}$, 
and taking into account that certain points are counted
(uniformly) finitely many times in 
the integral, we eventually obtain
$$
\|p^{1/2}\,\nabla\chi_R\,\nabla f\|^2_{L^2(\mathbb R^n)}\ \leq\
c_3\ \int_{B_{R+1}\setminus B_{R}}\ p\ (|\Delta f|^2+ |f|^2), 
$$
where $c_3$ is uniform in $R$.
By the dominated convergence theorem again, this integral 
goes to zero as $R$ goes to infinity. So, we finally obtain
$$
\lim_{R\ra \infty}\ \|p^{1/2}\ (\Delta f-\Delta f_R))\|_{L^2(\mathbb R^n)}
\ =\ 0.
$$

Since $f_R$ belongs to $H^2_c(\mathbb R^n)$,
the set of compactly supported functions in $H^2$,
we compute
$$
\nabla\cdot\left( p\ \bar f_R\ \nabla f_R\right)\ =\
p\ \bar f_R\ \Delta f_R\ +\ |\nabla f_R|^2\ p\ +\
\bar f_R\ \nabla p\ \nabla f_R.  
$$
Since $f_R$ has compact support, Stokes Theorem and our
hypotheses on $\nabla p$ show that
\bea\nonumber\hspace{-9mm}
\int p\,|f_R|^2 &=&\left|\int\ \bar f_R\ \Delta f_R\ p\ +\
\int\  \bar f_R\ \nabla p\ \nabla f_R \right|
\\ \nonumber\hspace{-9mm}
&\leq &
\left(\int |\Delta f_R|^2 p \right)^{1/2}
\left( \int |f_R|^2 p \right)^{1/2}
+\ 2\,C\,\left( \int |f_R|^2p\right)^{1/2}
\left(\int |\nabla f_R|^2p \right)^{1/2},
\eea
or, in other words,
$$
\|\nabla f_R\|_p^2\ \leq\ \|\Delta f_R\|_w\ \| f_R\|_w\ +\
2\,C\ \|\nabla f_R\|_w\ \| f_R\|_w.
$$
This estimate implies (\ref{ineq}) for $f_R$. The right hand side
of that estimate has a finite limit as $R\ra\infty$
with $f$ in place of $f_R$ on the right hand side. Since 
$$
\int_{B_R}\ p\ |\nabla f|^2\ \leq\ \int\  p\ |\nabla f_R|^2,
$$
we deduce that $p\ |\nabla f|^2\in L^1(\mathbb R^n)$
and satisfies (\ref{ineq}).
\hfill\ep

\vskip 5mm \noindent
{\bf Proof of Proposition \ref{dalpha}}\quad
We use the following Paley--Wiener theorem, Theorem IX.13 of \cite{rs2}:

\vskip 3mm\noindent
{\em Let $f\in L^2(\R^n)$. Then $e^{a|x|}f\in L^2(\R^n)$
for all $a<a'$ if and only if
$\widehat{f}$ has an analytic continuation to the set
$\{ p\,:\,|\Im p|<a'\} $ with
the property that for each $t\in \R^n$ with $|t|<a'$,
$\widehat{f}(\cdot +it)\in L^2(\R^n)$,
and for any $a<a'$,
$\ \sup_{|t|\leq a'}\ \|\widehat{f}(\cdot +it) \|_2<\infty$.}

\vskip 3mm\noindent
We refer to the conditions on $\widehat{f}$ in this theorem as 
``the Paley--Wiener conditions.''

Since $e^{a|x|}\,\ffi\in L^2(\R^n)$ is equivalent to
$\ffi\in D(e^{a\bra x \ket})$, 
Proposition \ref{eigv} shows that $\widehat{\ffi}$
is analytic everywhere and satisfies the 
Paley--Wiener conditions.
The functions $p\mapsto p_j\,\widehat{\ffi}(p)$ and
$p\mapsto\sum_j\,p_j^2\,\widehat{\ffi}(p)$
also satisfy these conditions. 

As a preliminary remark, we note that for any fixed
$t\in \R^n$, there exist $K(t)>\widetilde{K}(t)>0$
and $R(t)>0$, such that if $p\in\R^n$ satisfies 
$\sum_j\,p_j^2\geq R(t)$, then
\be\label{sqp}
\widetilde{K}(t)\ \sum_{j=1}^{n}\ p_j^2\ \leq\
\left|\,\sum_{j=1}^{n}\ (p_j+it_j)^2\,\right|
\ \leq\ K(t)\ \sum_{j=1}^{n}\ p_j^2.
\ee
So, if $B_{R}$ is a ball of radius $R$ with center
at the origin,
$-\widehat{\Delta \ffi}$ satisfies 
\be\label{init}
\int_{\r^n\setminus B_{R(t)}} \,
\left( \sum_{j=1}^{n} p_j^2 \right)^2\,
|\hat \ffi(p+it)|^2 \ dp < \infty,
\ee 
uniformly for $t$ in compact sets of $\R^n$.

We now start an induction on the length $|\alpha|$
of the multi-index $\alpha$ in $D^\alpha \ffi$.
We first show that $p\mapsto p_j\,p_k\,\widehat{\ffi}(p)$ 
satisfies the Paley--Wiener conditions
for any $j,k\in \{1,\cdots, n\}$. 
Note that we only need to prove estimates for large values
of the $|p_j|$'s. Also,
note that if $ \sum_{j=1}^{n} p_j^2\geq R(t)>1$,
there exists a constant  $C(t)>0$ such that 
\be\label{quad}
|\,(p_j+it_j)\,(p_k+it_k)\,|\ \leq\
C(t)\ \sum_{j=0}^n\ p_j^2,
\ee 
Therefore, (\ref{init}) implies that 
\bea
&&\int_{\r^n\setminus B_{R(t)}}\
|\,(p_j+it_j)\,(p_k+it_k)\,|^2\
|\widehat{\ffi}(p+it)|^2\ dp
\nonumber\\ \nonumber
&\leq& C^2(t)\
\int_{\r^n\setminus B_{R(t)}}\
\left(\,\sum_{j=1}^{n}\ p_j^2\,\right)^2\
|\widehat{\ffi}(p+it)|^2\ dp
\\ \nonumber
&<& \infty,
\eea
uniformly for $t$ in compact sets of $\R^n$.
Hence,
$\partial_{x_j}\partial_{x_k}\ffi\in D(e^{a\bra x\ket})$
for any $a>0$.

We next turn to third order derivatives.
Consider the derivative of
$-\Delta\ffi +(V-{\cal E})\ffi=0$.
For any $j\in\{\,1,\,\cdots,\,n\}$, 
$$
\partial_{x_j}\,\Delta\ffi\ =\
(\partial_{x_j}V)\ffi\ +\ (V-{\cal E})\ \partial_{x_j}\ffi.
$$
Since $V$ is a polynomial,
Proposition (\ref{eigv}) shows that 
$\partial_{x_j}\,\Delta\ffi\in D(e^{a\bra x\ket})$,
for any $a>0$. Thus,
the function
$p\mapsto p_j\ (\sum_{j=0}^n\ p_j^2)\
\widehat{\ffi}(p)$ satisfies the Paley--Wiener conditions. 

Consider now any triple of indices $j,k,l$. For 
$\sum_{j=0}^n\ p_j^2\ \geq\ R(t)$, we have
$$
|\,(p_j+it_j)\,(p_k+it_k)\,(p_l+it_l)\,|\ \leq\
C(t)\ |p_j+it_j|\ \sum_{j=0}^n\ p_j^2.
$$
Hence, using this estimate with (\ref{init}), we deduce that 
\bea
&&\int_{\r^n\setminus B_{R(t)}}\
|\,(p_j+it_j)\,(p_k+it_k)\,(p_l+it_l)\,|^2\
|\widehat{\ffi}(p+it)|^2\ dp
\nonumber\\  \nonumber
&\le&C^2(t)\ \int_{\r^n\setminus B_{R(t)}}\
|p_j+it_j|^2 
\left(\,\sum_{j=1}^{n}\ p_j^2\,\right)^2\
|\widehat{\ffi}(p+it)|^2\ dp
\\  \nonumber
&<&\infty,
\eea
uniformly for $t$ in compact sets of $\R^n$.
Therefore, the Paley--Wiener Theorem asserts that 
$\partial_{x_j}\,\partial_{x_k}\,\partial_{x_l}\,\ffi
\in  D(e^{a\bra x \ket})$, for any $a>0$.

We now proceed by assuming $D^\beta\ffi\in D(e^{a\bra x \ket})$,
for any $a>0$ and any $\beta$, such that $|\beta|\leq m$.
Let $\alpha$ have $|\alpha|=m+1$.
Let $\tilde\alpha$ be any multi-index of length $m-1$.
Differentiating the eigenvalue equation again, 
Leibniz's formula yields
\be\label{induc}
D^{\tilde\alpha}\Delta\ffi\ =\
\sum_{0\leq\gamma\leq\tilde\alpha}\
C^{\tilde\alpha}_\gamma\ \left(\,D^{\tilde\alpha-\gamma}
(V-{\cal E})\,\right)\ D^\gamma \ffi ,
\ee
where the $C^{\tilde\alpha}_\gamma$ are multinomial coefficients.
The induction hypothesis and the assumption that $V$ is a polynomial
show that $D^{\tilde\alpha}\ \Delta\ffi\in L^2(\R^n)$.
Therefore,
$p\mapsto p^{\tilde\alpha}\ (\sum_{j=1}^{n}\ p_j^2)\
\widehat{\ffi}(p)$
satisfies the Paley--Wiener conditions.
In $\alpha$, there are two indices, 
$\alpha_j$ and $\alpha_k$, not necessarily distinct,
which are larger or equal to one, such that
we can write
\bea\hspace{-9mm}
&&(p+it)^\alpha
\\[2mm] \nonumber \hspace{-9mm}            
&=&
(p_1+it_1)^{\alpha_1}\cdots(p_j+it_j)^{\alpha_j-1}\cdots 
(p_k+it_k)^{\alpha_k-1}\cdots(p_n+it_n)^{\alpha_n}\
(p_j+it_j)\ (p_k+it_k).
\eea
Estimating the absolute value of the last two factors by
(\ref{quad}) and using that\\
$\tilde \alpha=(\alpha_1,\cdots,\alpha_j-1,\cdots,
\alpha_k-1,\cdots,\alpha_n)$
has length $m-1$, we see that 
$p\mapsto p^\alpha\ \widehat{\ffi}(p)$
satisfies the Paley--Wiener conditions. Hence,
$ D^{\alpha}\ffi\in D(e^{a\bra x \ket})$
for any $a>0$.\hfill\ep

\end{document}